\documentclass[
,english
,authoryear
,A4
]
{elsarticle}
\usepackage[T1]{fontenc}
\usepackage[utf8]{inputenc}
\usepackage{color}
\usepackage{array}
\usepackage{graphicx}

\makeatletter

\providecommand{\tabularnewline}{\\}

\journal{Ecological Modelling}

\usepackage{array}
\usepackage{graphicx}
\usepackage{url}
\usepackage{breakurl}
\usepackage[breaklinks]{hyperref}

\makeatother

\usepackage{babel}
\begin{document}

\title{Simulations of populations of \textit{Sapajus robustus} in a fragmented 
landscape}

\author[biomas]{F. Keesen\corref{cor1}}

\ead{fbkeesen@gmail.com}

\author[defis]{A. Castro e Silva\corref{cor2}}

\ead{alcides@iceb.ufop.br}

\author[ufrg]{E. Arashiro}

\ead{earashiro@furg.br}

\author[defis]{C. F. S. Pinheiro}

\ead{cfelipespinheiro@gmail.com}



\cortext[cor1]{Corresponding author}

\cortext[cor2]{Principal corresponding author}

\address[biomas]{Programa de Pós Graduação em Ecologia de Biomas Tropicais, ICEB, Universidade
Federal de Ouro Preto, Campus Universitário s/n, Ouro Preto - MG, 35400-000, Brazil }

\address[defis]{Grupo Ciência da Complexidade, Universidade Federal de Ouro Preto, Departamento de Física, Campus
Universitário s/n, Ouro Preto - MG, 35400-000, Brazil}

\address[ufrg]{Instituto de Matemática, Estatística e Física, Universidade Federal do Rio Grande, Campus
Carreiros, Av. Itália, km 8, 96203-900, Brasil}


\begin{abstract}
The study of populations subject to the phenomenon of loss and fragmentation
of habitat, transforming continuous areas into small ones, usually
surrounded by anthropogenic matrices, has been the focus of many researches
within the scope of conservation. The objective of this study was
to develop a computer model by introducing modifications to the renowned
Penna Model for biological aging, in order to evaluate the behavior
of populations subjected to the effects of fragmented environments.
As an object of study, it was used biological data of the robust tufted
capuchin \textit{(Sapajus robustus)}, an endangered primate species whose geographical
distribution within the Atlantic Rain Forest is part of the backdrop of
intense habitat fragmentation. The simulations showed the expected behaviour based on the 
three main aspects that affects populations under intense habitat fragmentation: 
the population density, area and conformation of the fragments and deleterious effects due the 
low genetic variability in small and isolated populations. 
The model showed itself suitable to describe changes in viability 
and population dynamics of the species crested capuchin considering critical levels 
of survival in a fragmented environment and also, actions in order to preserve the species 
should be focused not only on increasing available area but also in dispersion dynamics. 

\end{abstract}


\maketitle

\section{Introduction}

Fragmentation and subsequent habitat loss is considered one of the
major causes of biodiversity depletion in natural 
environments~\citep{Fahrig2003,Gibson2011,Vitousek-1997}.
Effects of habitat fragmentation on populations will vary depending on species 
with different
life story 
strategy~\citep{Robinson1995,Ewers2006,Ferraz2007,Koprowski2005,Srinivasaiah2012}.
Features such as large body size \citep{Gehring2003,Lomolino2007},
specialized ecological requirements \citep{Dale1994} and extensive home 
range \citep{Crooks2002,Gerber2012} resulted in lowering the viability
of populations subjected to loss and habitat fragmentation. Besides
these effects take a role on entire structure of the community, such as changing
patterns of richness and abundance depending on the size and degree
of isolation of the fragments \citep{Laurance1997,Laurance2002,Laurance1998}.

Some species of vertebrates, such as primates, are especially sensitive
to loss and fragmentation of habitat. A recently published study showed
that the diversity and abundance of primates was directly proportional
to the size of the reserves \citep{Pontes2012}. The work of 
\citet{Chiarello2001}
observed that the population density of three species of primates
endemic to the Atlantic rainforest was directly correlated with the size
of the fragments studied, leading to the conclusion that the species population 
viability may be compromised in the
medium and long term in small fragments. \citet{Tuxtlas2007}
observed differences in diet, living area and population size of 
\textit{Alouatta
palliata} in fragmented areas of Mexico .

For the Atlantic rainforest biome, only 8.5\% of its original area still 
resists and is
heavily fragmented (SOS Mata Atlantica, 2013\nocite{SOSMataAtlantica;INPE2013}),
resulting in a great  number of species considered endangered of extinction. 
Nearly two thirds
of the 26 endangered species of primates in Brazil take place in the Atlantic
Forest \citep{MMA2014}. Concerning Sapajus, formerly
genus {\it Cebus} \citep{Alfaro2012}, three out of the four endemic species
of the Atlantic rainforest are considered endangered (\textit{S. xanthosternos},
\textit{S. flavius} and \textit{S. robustus}). From this, {\it Sapajus 
robustus} is among
the 10 most endangered primates on the Atlantic rainforest, being
the third most endangered of the Cebidae in Brazil
. Probably abundant in the past, the main threats to the populations
of this species are habitat destruction and intense hunting pressure
\citep{IUCN2015, MMA2014}.

In order to simulate the population dynamics of the species of primate
\textit{Sapajus robustus} subject to the effects of habitat fragmentation,
a computational model, whose bases were structured from acknowledged
Penna Model for biological aging \citep{Penna1995} was developed.
The Penna model is based on the ``Mutation Accumulation''\citep{MEDAWAR1952} 
theory, which states that
the pressure of natural selections decreases with age. That way,
bad mutations or deleterious alleles would be harmless in early stages of life 
when the selection is strong, however, in old ages, when
selection is weak, the sum of the effects of all mutations could cause 
dangerous outcomes in the organism. Only the
individuals that bypass this first stage (the strong selection) and reach the 
sexual maturity are able to transmit their genes.
Those genes start to accumulate in the population by genetic drift, leading 
to age evolution.

In this work we have developed a new version of sexual 
Penna model on a lattice with the introduction of a ``identification genome''
to take into account the effects of inbreeding, also simulating
different space configurations. It was named here as ``Fragmented Penna Model''.

This paper is organized as follows: Section 2: Modeling context; Section 3: 
Results and discussions;
Session 4: Conclusion and Future Perspectives.

\section{Modeling context}

\subsection{Penna model of aging - sexual version}

The Penna model \citep{Penna1995a} is a computer model based in accumulation of 
mutation theory. This model has successfully explained the
semelparous senescence of Pacific salmon\citep{PhysRevE.52.R3309}, the control 
mechanism of menopause in parental care on sexual populations 
\citep{MossdeOliveira1999},
the process of sympatric speciation\citep{Oliveira2003}, the effects of 
temperature over population dynamics\citep{DeOliveira2008},
the emergence of chaotic 
behaviour\citep{bernardes1998simulation,castro2001analysis} and recent 
simulations
have shown some results regarding population stability and carrying capacity 
\citep{Pinol2011a}. The first works with sexual version of Penna 
model was carried by \citep{BernardesAT1, Stauffer1996}. 
\citet{sousa2000simulating} simulated inbreeding depression through mutation 
accumulation theory.
These authors tested, among other things, the effect of the influence of
environmental conditions on reproduction and its consequences in population 
aging -
longevity and catastrophic senescence. More recently, the Penna sexual model 
has been used not only as
a mathematical perspective, but also considering simulation applied to the 
biological insect control\citep{deSouza2009756}.
Another adaptations using species characteristics, environmental factors and 
others 
were performed as well\citep{Magdon1999182,kim2012modified} .

In the sexual version, each individual
is represented by two ``alleles'' or {\it chronological genomes} given by 
``bit'' strings formed by two sequences of ``0s'' and ``1s''.
These sequences have a length of a computer word (usually 32 or 64 bits) that 
represents the organism's life span. The bits in
these words are time ordered in ``years'', and a bit ``1'' in a given position 
represents a bad mutation that will lead to a disease in
that related year. The disease will be expressed taking into account if the 
``1s'' appears in both alleles (homozygous) or if the bad
mutation is dominant. The diseases are accumulative, and the organism dies at 
the age when $T$ diseases are expressed.
The whole population is aged structured and when sexual maturity is reached, 
couples are formed randomly among males and females with
age greater than $R$. The reproduction is done with crossing-over recombination 
of the parents words, and $m$ mutations are introduced
in random positions of the $B$ offsprings {\it chronological genomes}. The 
gender of the descendants is chosen between male or female
with 50\% chance. In order to simulate a carrying capacity of the environment, 
$K$, preventing an unrealistic growth of the population,
each individual can die with a probability given by the Verhulst factor, namely 
$V=N(t)/K$, where $N(t)$ is the number of individuals
at time.

\subsection{Biological parameters of the model}

The present model was built considering the inherent biological parameters
{\it Sapajus robustus} species from published studies (Table 1).
 Given the
taxonomic conflict around the old genus {\it Cebus}\footnote{\citet{JUNIOR2001} 
elucidated the taxonomy of the genus Cebus, \textit{C.
robustus} revalidated as a species. More recently the species was
separated into another genus: Sapajus \citep{Alfaro2012}.}, this species was 
considered for many years as a subspecies of \textit{Cebus
apella} \citep{Torres-de-assuncao-1983} or, as a subspecies of \textit{C.
nigritus}. That brings a lack of unique ecological data for the species.
Thus, information that refer to genus, prioritizing those alluding
to \textit{Sapajus nigritus} were used. This species is phylogenetically
close to \textit{S. robustus}, and has its geographical distribution in the
Atlantic Forest Biome \citep{Alfaro2012,rylands2005} 

The species is distributed geographically in an area of intense interference
on landscape by human activities, with extensive fragmented areas,
and subject to various uses and occupations of the soil, which places
it in a situation of greater vulnerability than \textit{S. nigritus}.

All the information on the ecology and reproduction of the species,
considered essential to feeding and dynamics of the model (Table 1),
were then raised. Every significant aspect to the understanding of
social structure and behavioral information were also assessed for
information and indirect validation of the model.

\begin{table}

\caption{Relevant aspects of genre \textit{Cebus/Sapajus} for the model.}

\begin{tabular}{>{\centering}m{0.2\columnwidth}>{\centering}m{0.1\columnwidth}>{\centering}m{0.5\columnwidth}}
\hline
{\small{}Atribute} & {\small{}Value} & {\small{}References}\tabularnewline
\hline
{\small{}Dispersion age (male) (average)} & {\small{}5.7 years} & 
{\small{} \citep{DiBitetti2001,Janson2012}}
\tabularnewline
{\small{}Sexual rate} & {\small{}01:01:00} & 
{\small{} \citep{Fragaszy2004}}\tabularnewline
{\small{}Longevity (average)} & {\small{}32 years} & 
{\small{} \citep{DiBitetti2001,Janson2012}}\tabularnewline
{\small{}Male sexual maturity } & {\small{}5 - 7 years} & 
{\small{} \citep{DiBitetti2001,Janson2012}}\tabularnewline
{\small{}Female sexual maturity} & {\small{}5 - 8 years} & 
{\small{} \citep{DiBitetti2001,Fragaszy2004}}\tabularnewline
{\small{}Gestation length (average)} & {\small{}6 months} & 
{\small{} \citep{DiBitetti2001,Fragaszy2004,Janson2012}}\tabularnewline
{\small{}Fertility rate} & {\small{}1} & 
{\small{} \citep{DiBitetti2001,Fragaszy2004,Janson2012}}\tabularnewline
\hline
\end{tabular}
\end{table}

\subsection{Fragmented Penna Model}

In order to simulate the \textit{Sapajus robustus} behaviour in fragmented 
environment, some adaptations were made in the Penna model.
The first is the introduction correct set of parameters (Table 1). In this
case, $R=7$, lifespan=$32$, $B=1$ and $T=5$.
The second one is the environment. In the original Penna model there is no 
definition of the environment, or ``space'' where the dynamics
takes place. There is just one environmental parameter, the carrying capacity 
$K$.

In the fragmented model, the simulations are done in square lattices of size 
$L=N \times N$, with one individual for cell. In the initial
condition of the dynamic ($t=0$) the cells are filled randomly with equal 
amount of males and females, following a density $\rho(0)$.
Each cell is labeled according to the sexual status of its individual at 
present time: (-1) for young males,
(-2) sexually mature males, (1) young females, (2) sexually mature females, (3) 
pregnants and (4) parental care. 
It is important to note that, even if we do not take into account dispersion 
and sexual preferences, males and females have 
different reproductive stages during its lifetime. Pregnant females and females 
in parentar care are unable to reproduce while 
those states does not occur in males. The use of so many states are, in this 
case, necessary to perform a more realistic simulation 
of this species reprodutctive behaviour. 

 The simulation can be summarized as follows.
Each time step is considered one year, in this period there are the steps of 
migration, reproduction, birth and death. 
At each time step first occurs the displacement of individuals. The sum of 
empty sites on
Moore neighborhood\footnote{The Moore neighborhood is defined on a 
two-dimensional square lattice and is composed of a central cell and the eight 
cells which surround it.} of each individual comprises its space accessible on 
that
dispersion round. After that, the accessible sites are drawn in random order to 
be occupied by
individuals in their neighborhood. When there are more individuals than empty 
sites, the individuals
are chosen randomly. Males are allowed to perform larger displacements than 
females, since they can cross the border of their territory, under the 
probability of dispersion $\phi$.

After the dispersion round, it is time for matching and reproduction, given by 
the following rules:
\begin {enumerate}
\item Male and female must be sexually mature, that is, their age must be 
greater than $R$.
\item Females in parental care state (4) are unable to reproduce during 2 time 
steps.
\item Females choose males in her Moore neighborhood.
\item One couple will generate an offspring only if there was an empty cell 
available on neighborhood to be occupied by these offspring.
\end{enumerate}
The displacement of the offspring is leaded by the mother. When the mother 
moves one cell, the offspring takes
the previous mother position, following her during 2 time steps, never leaving 
the mother's neighborhood.
Finally, death comes by disease, for the individuals that reach the threshold 
$T=5$.

The last adaptation is regarding to the length of the computer word used to 
model each individual. In this work,
instead using words of 32 bits, each individual is represented by two words of 
64 bits. The first 32 positions are
``alleles'' inherited from parents during sexual reproduction and used on the 
aging dynamics.
The last 32 positions are used to identify the groups in fragmented areas, and 
called ``identification genome''. The
presence of this genome introduce the last modification in this version of 
Penna model. Apart from the random mutations
introduced in reproduction, was inspect how similar the ``identification 
genomes'' of the parents are. The 
``identification genomes'' are inherited from mothers, having no role in 
crossing over and reproduction. If the genomes
are similar (meaning the parents are closed relatives), another deleterious 
mutation is introduced. The number of mutations
inserted is proportional to the similarity as follows: if the similarity of the 
couple is in the range $[48,63]$ (meaning that they have at least one common 
grandparent), one mutation is added; if the similarity reaches $64$ (the
couple are brothers), the amount of mutations increase to $3$. The similarity 
of two individual is measured comparing
their ``identification genomes'' one by one, which result in $4$ combinations, 
and choose the greater one as result. 
The introduction of different mutation rates was done in order to simulate 
inbreeding. 

\subsection{Functionality}

To test the functionality of the model, the first analysis was performed
on the behavior of the population as a function of the initial
density ($\rho(0)$). Population at time zero consists of random
samples of $N(0)$ individuals with random bit-strings (both chronological
genome and identification strings), ages uniformly distributed between
1 and 7 years, and female or male sex equally likely. The density
at any time is computed as $\rho(t)=N(t)/L$. Thus, it was necessary
to assess what value of initial density would provide 
\textquotedbl{}chronological
genome\textquotedbl{} variability for the population, from the point
of temporary stabilization of the curve, were able to maintain viable
populations for the model simulations. The initial density values
ranged from 0.1 to 1.0 for three sizes lattice: L=$10\times10$ (100
sites or cells), L=$20\times20$ (400 sites or cells), and L=$30\times30$
(900 sites or cells) (See Results, subsection 3.1).

The second analysis is related to how the area affects the population
behavior. It was done by running the model with same initial condition
$\rho(0)=0.8$ for six different sizes of square lattices. In this case, the
effect of area is measured without considering possible interactions
between sub-lattices or the influence of variability of the 
\textquotedbl{}identification
genome\textquotedbl{} in the results.  (Subsection 3.2).

In the subsection 3.3, the same analysis was performed, but 
individuals, represented by dispersing males, were allowed to migrate from one 
territory
to another. Those fragmented territories are represented by square sub-lattices.

In the subsection 3.4 was carried an investigation about the effects
of geometric configuration in demographic behavior for each type of
sub-lattice with and without dispersion of individuals. Lattices with sizes
L = $100 \times 100$ were divided into subnets of equal areas, but with 
different
geometric configurations. The different subnets were here inserted to map the 
classic formation
of patches subject to the so-called ``edge effects'' in fragmented
environments \citep{Murcia1995}. They can also represent the demographic
effects caused by loss of effective area in population viability 
\citep{Fletcher2007a}.
The edge effect tends to decrease the amount and diversity of resources,
which may disrupt the balance between consumption and production that
affect the carrying capacity of the environment \citep{Fagan1999a,SAUNDERS1991}.
Besides directly influencing the mobility of organisms, altering the
flow of emigration and immigration \citep{Ries2004}.

Finally, in subsection 3.5 a study was developed about the effect that would 
have on 
population survival if the migration was
made by females instead of males, regarding individuals dispersion. To avoid 
the influence of the previously
observed area effects, it was used a fixed-size square lattice L=$20 \times 20$ 
subdivided into four $10 \times 10$ sub-lattices.

\section{Results and discussion}

\subsection{The effect of density in population}

\begin{figure}
\begin{centering}
\includegraphics[width=0.9\linewidth]{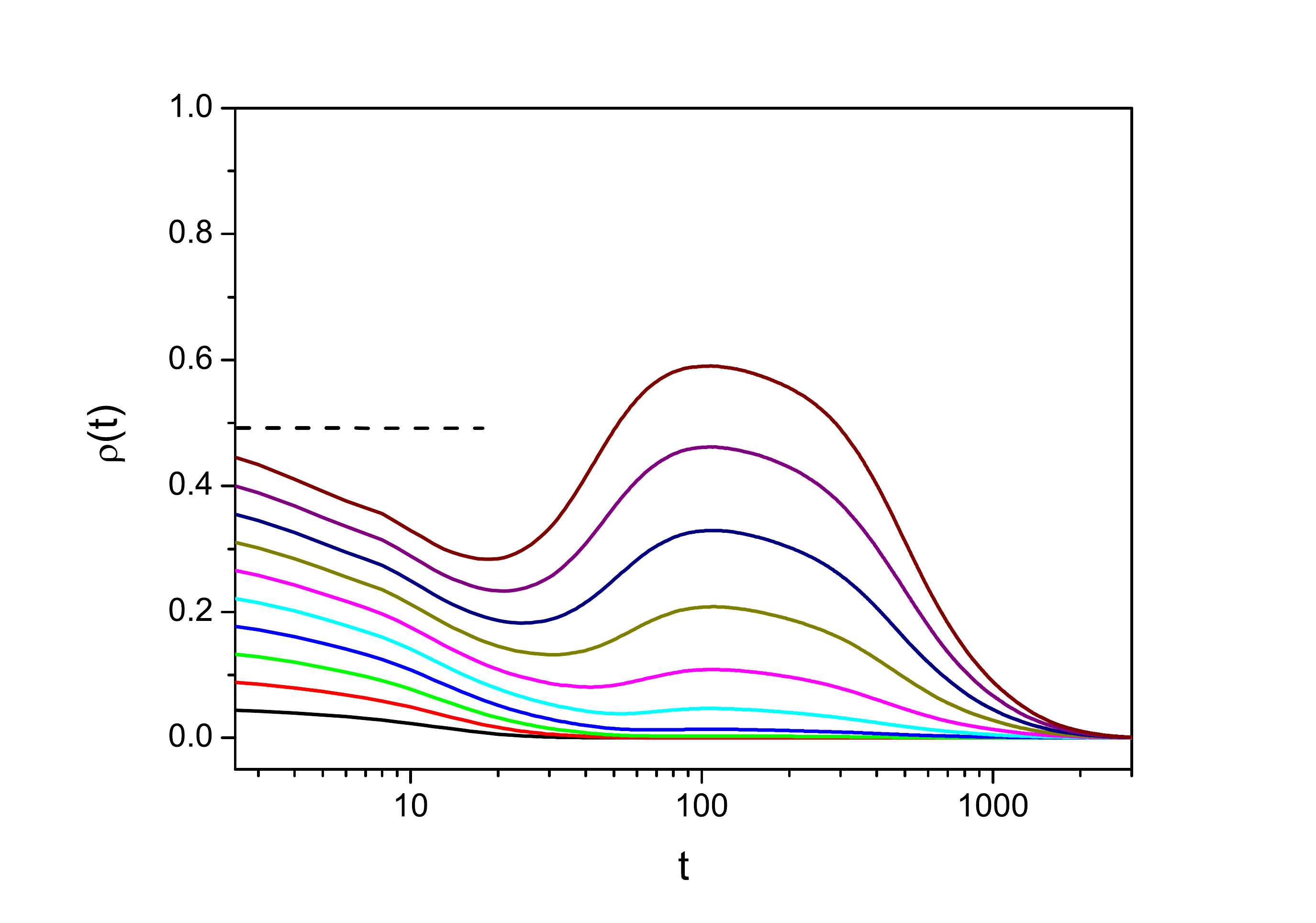}
\par\end{centering}

\caption{\label{fig:figure01}The influence of initial density on time evolution
of population densities for lattices of size $L=10\times10$. Averaged
over 20,000 samples. The horizontal bar indicates the genome selection
time period. The curves are arranged in decreasing order of $\rho(0)$, from 
$\rho(0)=1.0$ (uppermost) to
$\rho(0)=0.1$ (bottommost), the initial density spacing between the curves is 
$\Delta\rho(0)=0.1$.}
\end{figure}

For a lattice with L =  $10\times10$, considered small, any initial density
condition leads to the disappearance of the population without presenting
any stability period (Figure~\ref{fig:figure01}). However, the behavior
of the population becomes different as the initial density reaches
0.5.
For initial densities below 0.5, the population fell continuously
until the total disappearance of the population, i.e., $\rho(0)<0.5$
translates into a population on which the number of individuals who
have a viable \textquotedbl{}chronological genome\textquotedbl{} is
insignificant. The model eliminates individuals who fail to reach
reproductive age at the threshold $T$. Without a minimum number of
breeders, the population tends to quickly fade. It is important to
note that, due the huge simulation time steps,
the horizontal time axis was changed to logarithmic. Therefore, the initial 
density $\rho(0)$
cannot be taken as the intercept of the curves with $y$-axis. In this plot, the 
curves
are arranged in decreasing order of $\rho(0)$, starting from $\rho(0)=1.0$, the 
uppermost curve,
to $\rho(0)=0.1$, the bottommost one. The same
applies to all plots with logarithmic time axis.

Initial densities above 0.5 show a initial drop in population, corresponding
to the selection of viable chronological genome (dashed horizontal bar in
Figure~\ref{fig:figure01}), followed by population growth; these
growth rates are increasing functions of initial density values $\rho(0)$.
Because space is the limiting factor here, populations behave as confined
populations, where not even an increase of individuals selected in
time is able to overcome the effects of a small population and low
variability. It is possible that parentage factor leads to a
rapid population decline due to the increasing number of deleterious
mutations, analogous to the expected effects for small populations:
such as demographic stochasticity and genetic drift 
\citep{Young2000,ALLENDORF2013}.

\begin{figure}

\begin{centering}
\includegraphics[width=0.9\linewidth]{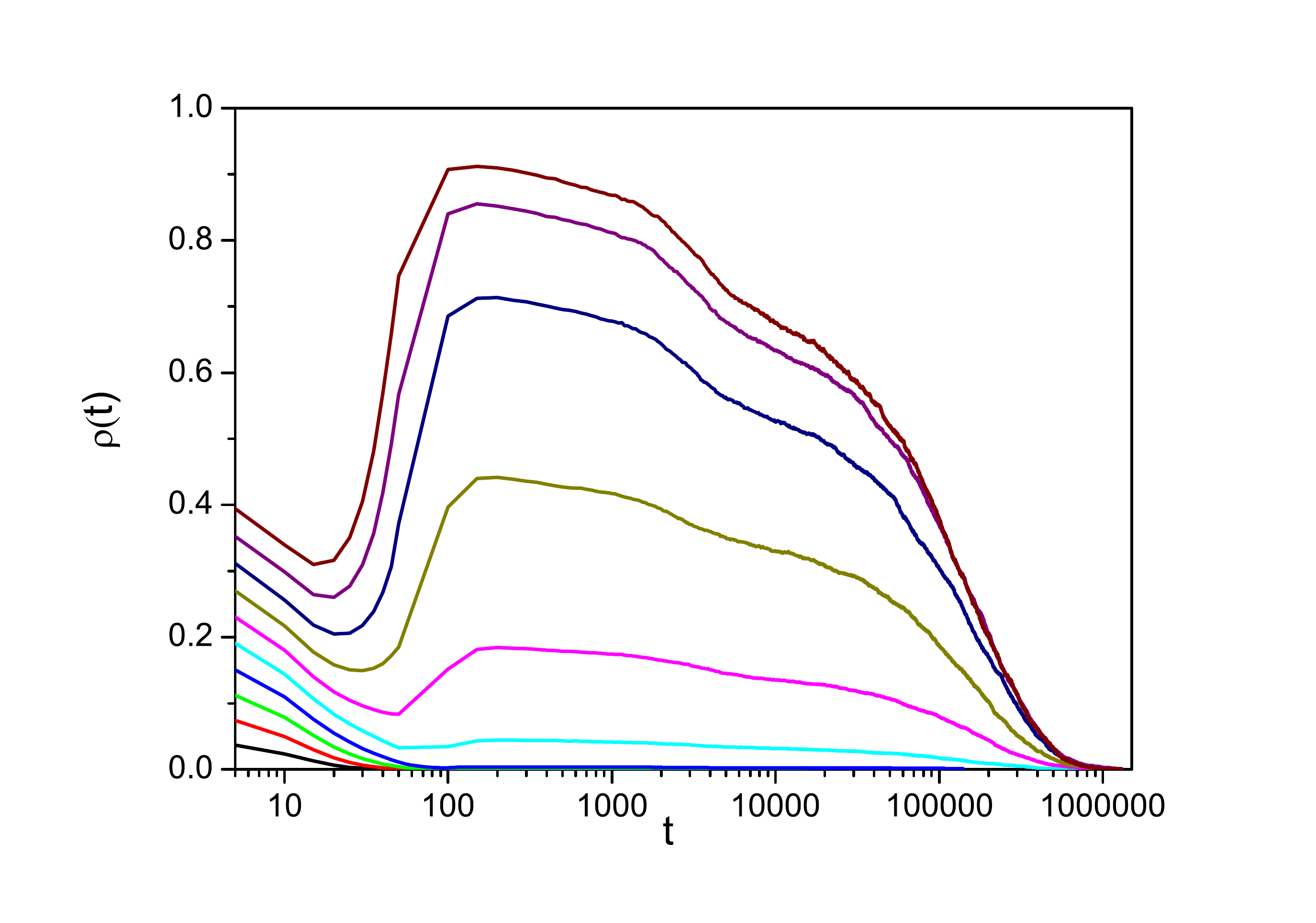}\\
\end{centering}

\caption{\label{fig:figure02}The influence of initial density on time evolution
of population densities for lattices of size L = $20\times20$. Averaged
over 10,000 samples. The curves are arranged in decreasing order of $\rho(0)$, 
from $\rho(0)=1.0$ (uppermost) to
$\rho(0)=0.1$ (bottommost), the initial density spacing between the curves is 
$\Delta\rho(0)=0.1$. 
A much longer persistence of population is found when compared to $10\times10$ 
lattices.}
\end{figure}

Similar outcomes were found for lattices of size L = $20\times20$ 
(Figure~\ref{fig:figure02}), with
the minimum viable initial density $\rho(0)=0.5$. Although, in this case, 
populations
now have a much longer persistence time, they eventually fade -- this
is something inherent to the Penna model. Also, larger values of $\rho(0)$
do not result here in longer persistence times, the influence of the
space limitation factor.

\begin{figure}

\begin{centering}
\includegraphics[width=0.9\linewidth]{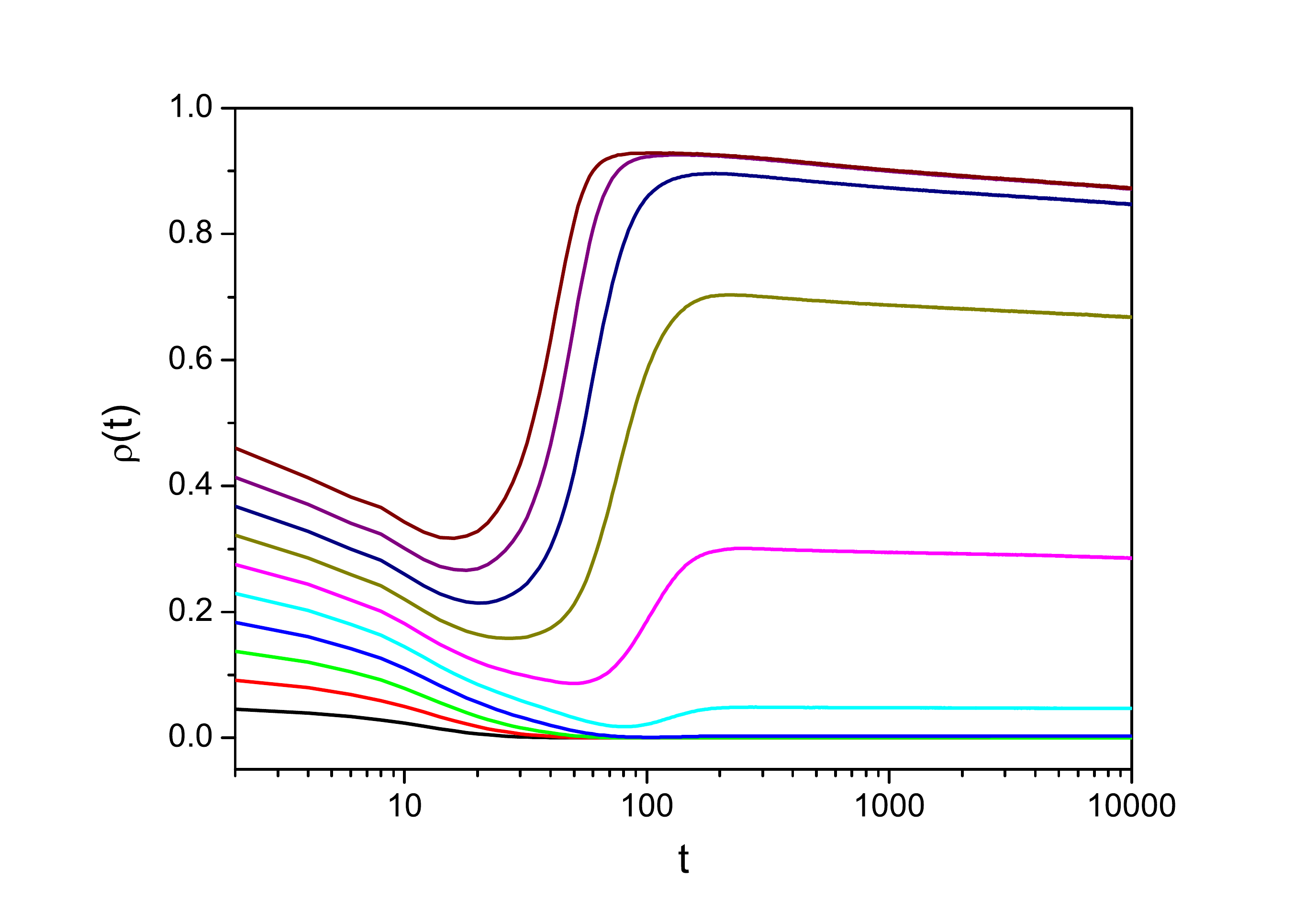}
\end{centering}

\caption{\label{fig:figure03}The influence of initial density on time evolution
of population densities for lattices of size L = $30\times30$. Averaged
over 1,000 samples. L = $30\times30$ lattices (bottom plot)
can be considered as large, since they can support long standing populations,
indicating ``genomic'' variety persistence. The curves are arranged in 
decreasing order of $\rho(0)$, from $\rho(0)=1.0$ (uppermost) to $\rho(0)=0.1$ 
(bottommost), the initial density spacing between the curves is 
$\Delta\rho(0)=0.1$.}
\end{figure}


In lattices of size L = $30\times30$, after the initial ``chronological
genome'' selection, the population recovers and remains at nearly
constant density (Figure~\ref{fig:figure03}). The extremely low rate of 
population decline, with
no predictable sudden drop even for very long simulation times \footnote{The 
simulation times were limited by computer available RAM.},
make L = $30\times30$ large-size lattices for the model. That is, to
a large and densely populated lattice, the effect of the accumulation
of deleterious mutations is not observed since a broader range of sexual
partners will be present. A discernible effect in space limitation
is the saturation of population density for $\rho(0)\geq0.8$. It
is possible to say that the population reaches a steady state for
an indefinitely long time, which means that the effect of deleterious
mutations does not reach the threshold to take the population to extinction.

Low initial density values lead the population to disappearance in
a short time, for all lattice sizes, and also below critical $\rho(0)$
values in a viable size areas (L=$20\times20$, $30\times30$). There
is a minimum $\rho(0)$ value 0.5, above which the population recovers
and remains viable. The increased area and density variation from
that minimum value will give the population higher survival rates,
but always controlled by the environment capacity of the support,
represented in this model by the space factor (available sites and
isolating areas). In nature the space variable can be represented
by the space itself. A good example would be a landscape of small
fragments without connectivity; and the effective space or lack of
proper resource: a large fragment in size, but composed of advanced
forest interleaved with dirty pasture, or an area of mono-specific
reforestation could represent a large area in spatial terms, but a
non-effective area in ecological requirements.

The model developed was able to exhibit two major aspects related to population
dynamics: the relationship between density-dependent 
growth\citep{Lotka1925,Volterra1926}
and the so-called minimum viable population \citep{Soule1985,Shaffer1981}.

The density-dependent population growth reveals that low densities
are not able to maintain viable populations, because they affect the
rate of population growth if the density reaches a critical value.
Low densities in nature represent a relationship between reproductive
fitness of the individuals of a species and the density of conspecifics
\citep{Allee1927}, lead to a decrease in the population growth
and may cause, for example, local extinctions. Several factors are
identified as causes of this effect, such as changing the values of sex
ratio \citep{Society1994}, susceptibility to predation
\citep{Gascoigne2004} and especially the demographic aspects, such
as active dispersal. For instance, demographic aspects can lead to the decrease
in population growth by loss of individuals \citep{Bonte2004,Young2000}
and genetic factors such as inbreeding depression, result from the
effects of genetic drift.

In this model, the concept of a minimum viable population was evidenced
by the existence of a minimum viable density $\rho(0) \sim 0.4$. Minimum
viable population or minimum number of individuals in an ideal area
refers to a minimum population capable of ensuring the persistence
of the population in a viable state for a given time interval \citep{Rai2003},
i.e. maintaining gene flow and genetic diversity \citep{ALLENDORF2013}.

\subsection{Effect of area in population behavior}

Based on the results of density observed for three lattice sizes,
$\rho(0)=0.8$ was adopted to test the model and analyze other
parameters. A density $0.8$ is able to support a sufficiently large
variability of chronological genomes so that initial density is not
a bias in the evolution of the model.

\begin{figure}
\begin{centering}
\includegraphics[width=0.9\columnwidth]{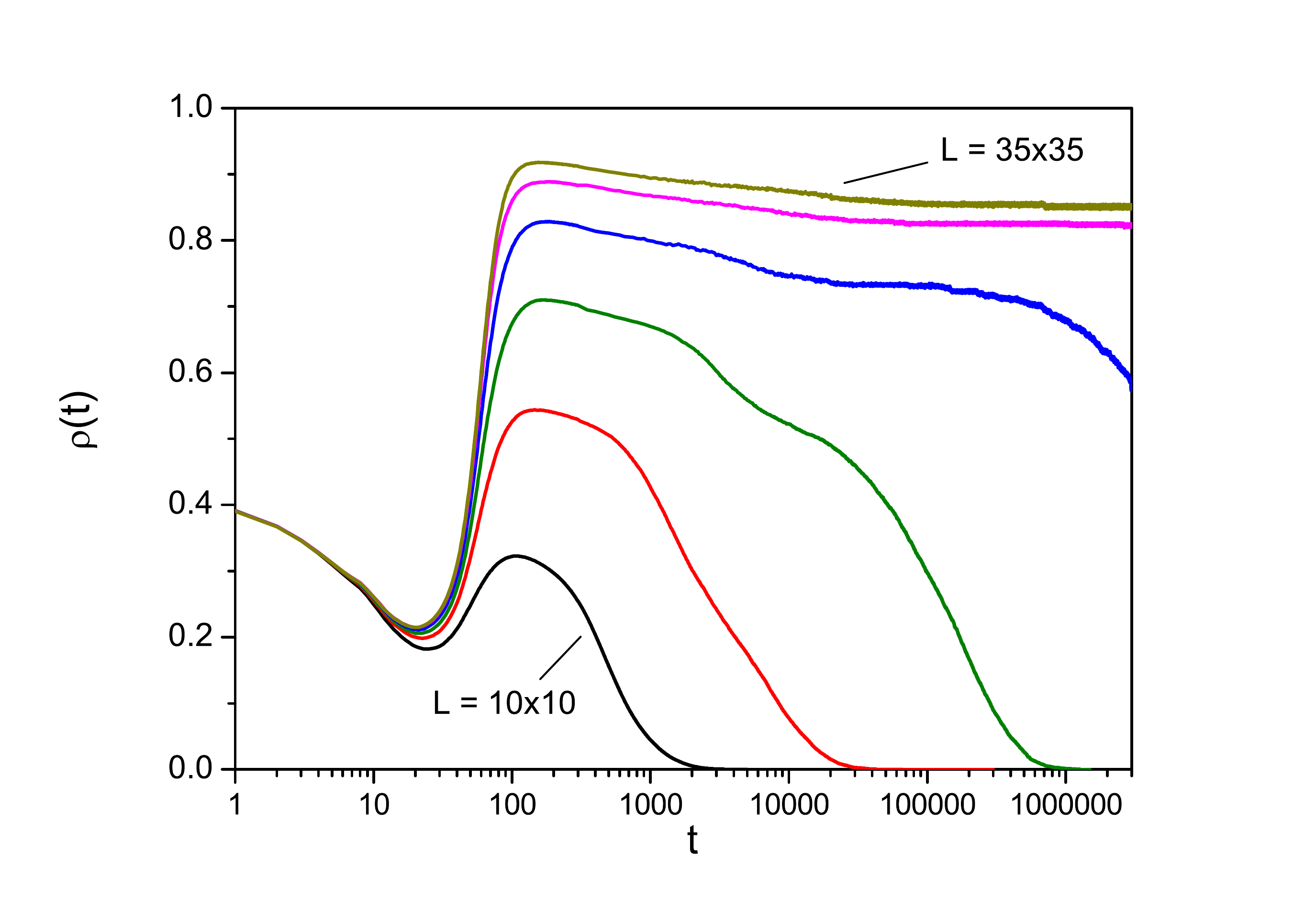}
\par\end{centering}

\caption{\label{fig:figure04}Effect of population area on viability. The
effects of area for various lattice sizes (see legends) with an initial
density $\rho(0)=0.8$ were evaluated. Averaged over 20,000 samples
for L=$10 \times 10$, 10,000 for L = $15 \times 15$, 5000 to L = $20 \times 
20$, $25 \times 25$
to 2000 and 600 to L = $30 \times 30$ and $35 \times 35$.}
\end{figure}

Six different sizes of square lattices were considered and results
can be found in Figure~\ref{fig:figure04}. Clearly the lattice
size (area) directly affects the viability of the population. Disregarding
the initial occupation effect, a population confined to a small
area tends to fade much faster than a population with the same rate
of occupancy in a large area. Another behavior extract from this graph
is that populations in a small lattice, although they can recover
from the chronological genome selection, they do not maintain stabilized
values, that is, the behavior is always downward to extinction. From
L=$20 \times20$ on, it is observed the existence of a period of population
stability, the length of which increases as lattice size increases,
reaching a size of $30 \times 30$, where the space effect ceases. 
L=$30\times30$ would
then be an area size capable of maintaining a viable population for
an indefinitely long period of time.

The effects of mutation accumulation seem to be diluted for large
lattices. Since in this case, more groups with different ``identification 
genomes'' are present and also space mobility is greater, the population
probably evolves with a weaker spatial correlation between siblings
and parents, decreasing the probability of inbreeding. This aspect
is further discussed ahead, when the inbreeding rate was analyzed in 
different scenarios.

Small lattices represent areas unable to maintain viable population
for a long time. Two recent studies have shown the effects of area
on the richness and abundance of primates: \citet{Boyle2010a} analyzed
the distribution and persistence of six species of primates found
in forest fragments, isolated from the year 1989 in the Brazilian
Amazon. They have found a direct correlation between the wealth of
primates and the size of the fragments. \citet{HARCOURT2005} conducted
a comprehensive analysis of the species-area relationship for tropical
forest primates in studies compiled for about 133 fragments, distributed
in 33 areas of the five continents. Data analysis indicated that the
wealth and the proportion of the wealth of primates decrease in an
almost linear relationship with the area of the fragment.

In this model the area showed an almost direct relation to population
density. Lack of resources limits growth, leading to the intrinsic
effects of small populations. On the other hand, a large area with few 
individuals
and no external inputs (no migration) will suffer almost the same
effect as a small, densely populated area: fewer individuals, fewer players,
changes in the reproductive rate and low variability (genetic diversity)
-- \textquotedbl{}identification genomes\textquotedbl{}. It is worth
noting, however, that in nature , this relationship may vary considering
intrinsic factors of the ecology of species. As example, the size of the
living area \citep{Mitani1979,Pearce2013}, body size \citep{Bennett2009}
and the degree of specialization with regard to ecological requirements
\citep{Chaves2012a,Dale1994,Gascon1999,Puttker2013} becoming something
more or less susceptible to the mentioned effects.

Studies for the genus {\it Cebus} showed that most species have large living
areas, although they are able to keep up with high densities in small
areas due to two main factors: the type of resources available in
the area and; the ability to exploit the resources seasonally 
\citep{Fragaszy2004}.
Although this model is not able to describe the individual contribution
of each resource factor that controls the population dynamics of the
species in the wild, the occupation of sites and parentage parameters
have represented, so far, the general behavior of the population from
critical levels of survival (a small, isolated area) to viable levels
(a large continuous area).

\subsection{The viability of population in square fragments}

\begin{figure}
\begin{centering}
\includegraphics[width=0.9\columnwidth]{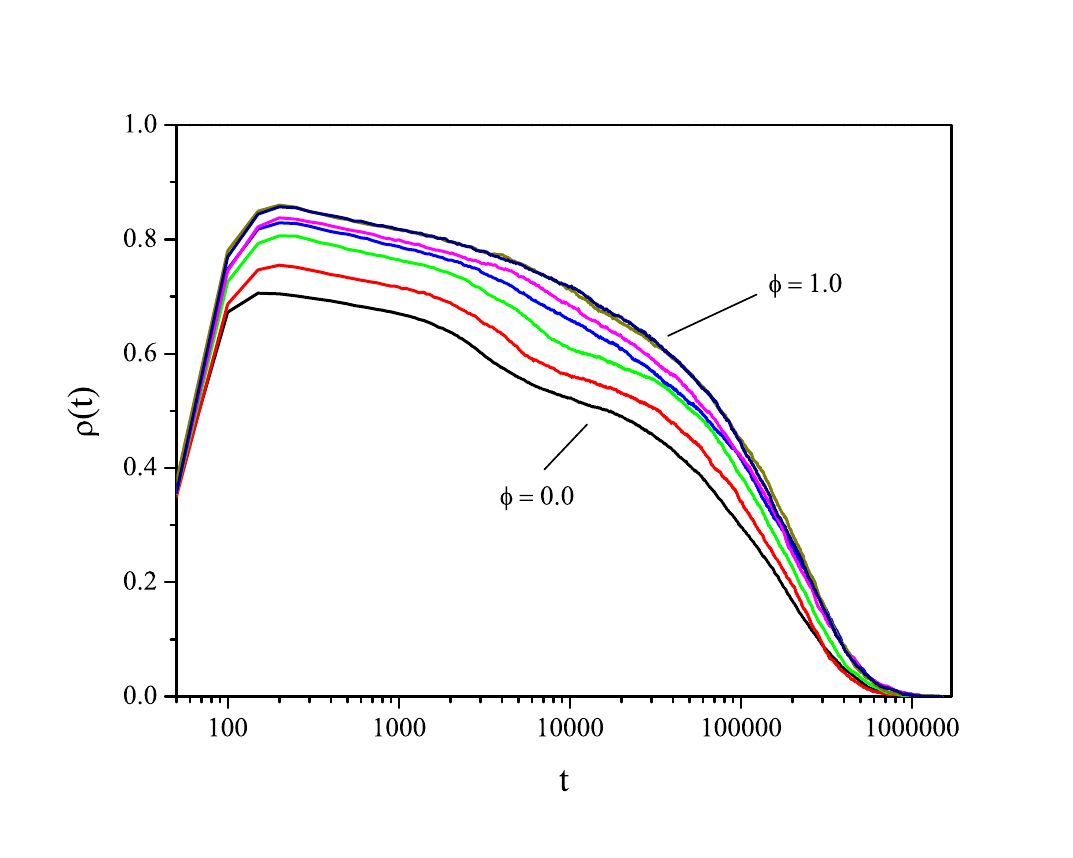}
\par\end{centering}

\caption{\label{fig:figure05}Effect of dispersion in a lattice size $L= 
40\times40$ subdivided into 
four sublattices of $L = 20\times20$ with an initial
density $\rho(0)=0.8$ . The curves, from bottom to top, are for 
dispersion probability $\phi$ = 0.0, 0.1, 0.2, 0.3, 0.6. 0.8 and 1.0.}

\end{figure}

Considering a lattice of intermediate size, considering 400 sites
for each sublattice (area), it is observed that from a (across the border)
dispersion rate of $\phi=0.2$ the effects of variability can modify
the behavior of the curve; although it will not generate increased
survival (Figure~\ref{fig:figure05}). Thus, this value was
defined as the cutoff, maintaining a dispersion of active probability
throughout life, from 7 years old, and for males only.

\begin{figure}
\begin{centering}
\includegraphics[width=0.9\columnwidth]{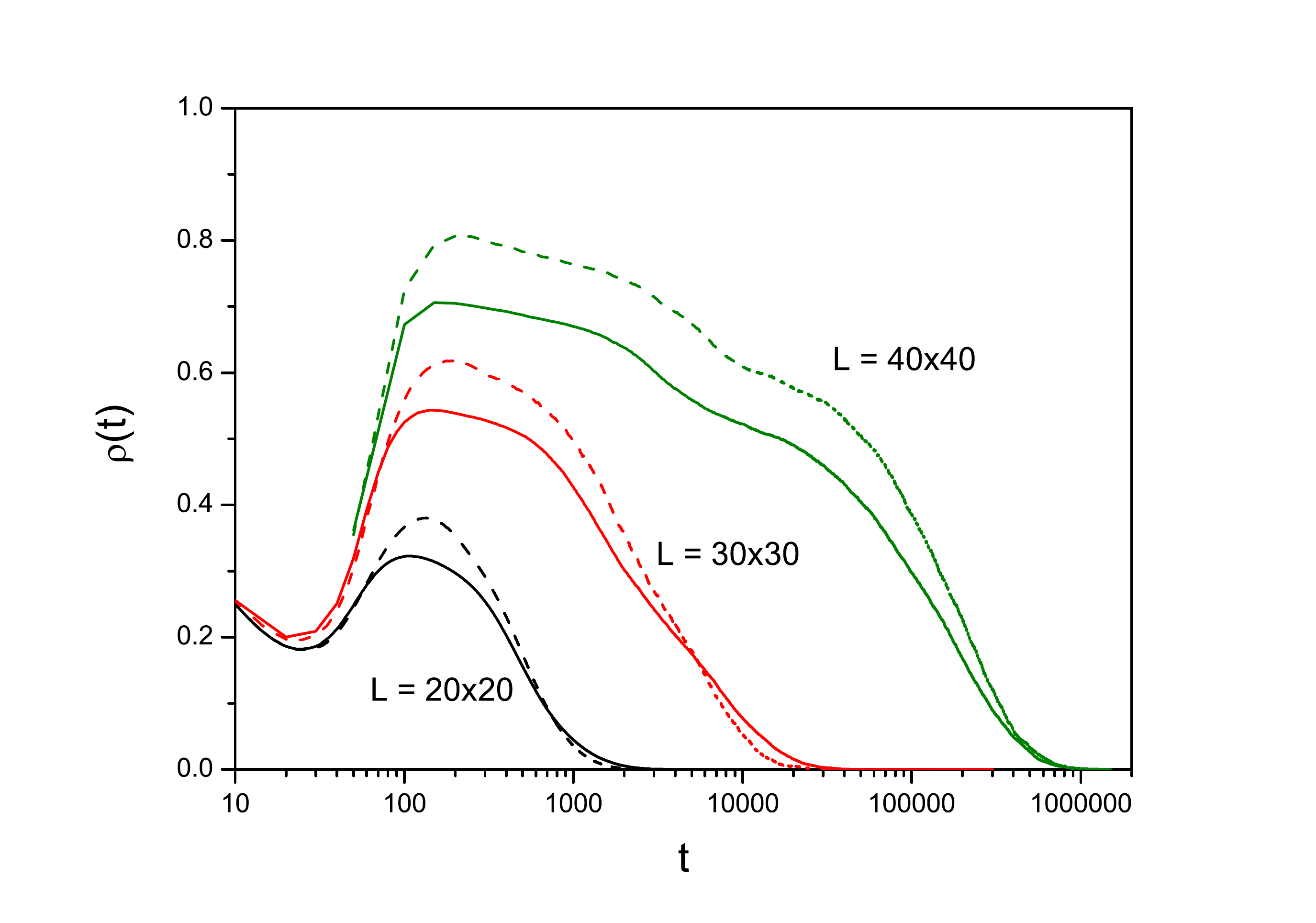}
\par\end{centering}

\caption{\label{fig:figure06}Density $\rho$ plotted as a function of time ($t$)
for three different square lattice size subdivided into four sublattices.
Effects with dispersion, $\phi(0)=0.2$, (dashed line) and without
dispersion (full line) were considered.  Initial density $\rho(0)=0.8$}
\end{figure}

When evaluating the effect of dispersion for different sizes of square
sublattice, the change in the behavior of demographic population seems
to be ineffective since even by increasing the size of lattices, i.e.
eliminating the effect of the area, the dispersion changes the rate
of population growth without, however, increasing the survival rate
(Figure~\ref{fig:figure06}). For small sublattices ($10 \times 10$
and $15 \times 15$) dispersing the population leads to a more rapid decrease,
making it clear again that the effects of density (initial density)
and the lattice size (size of the available area) reflects directly
on the viability of the population. The flow of reproductive partner
migrating from one lattice to another is insufficient to cause a visible
change since the number of individuals as a whole is insufficient
for the simple reproduction behavior. In this conformation, migration
may represent a larger loss for the origin territory than it is a
gain to the destination patch, which may be already irreversibly doomed.

The disappearance of the population with and without dispersion occurs
essentially at the same time step, representing long-term absence
of effective flow considering the behavior of the model 
(Figure~\ref{fig:figure06}).
The behavior of dispersion of viable sub-lattice (L = $20 \times 20$ ) seems
to work as a leakage flux of males between different social groups
within a continuous area. The flow occurs within the Hardy-Weinberg
Balance for males (constant allelic frequency), i.e., maintaining
the genetic diversity of the population. Thus, other factors such
as space, resource competition and predation will control and maintain
the population at a value of dynamic equilibrium \citep{Gotelli2008}.
The dispersion, for this lattice configuration, will not alter survival,
which is already controlled by the carrying capacity of the environment.
Noting that the disappearance of the entire population, even for viable
lattice sizes, is a consequence of the accumulation of deleterious
mutations, inherent to the original model itself.

\subsection{The viability of population in sub-lattices with different geometric
configurations}

\begin{figure}
\begin{centering}
\includegraphics[width=0.9\columnwidth]{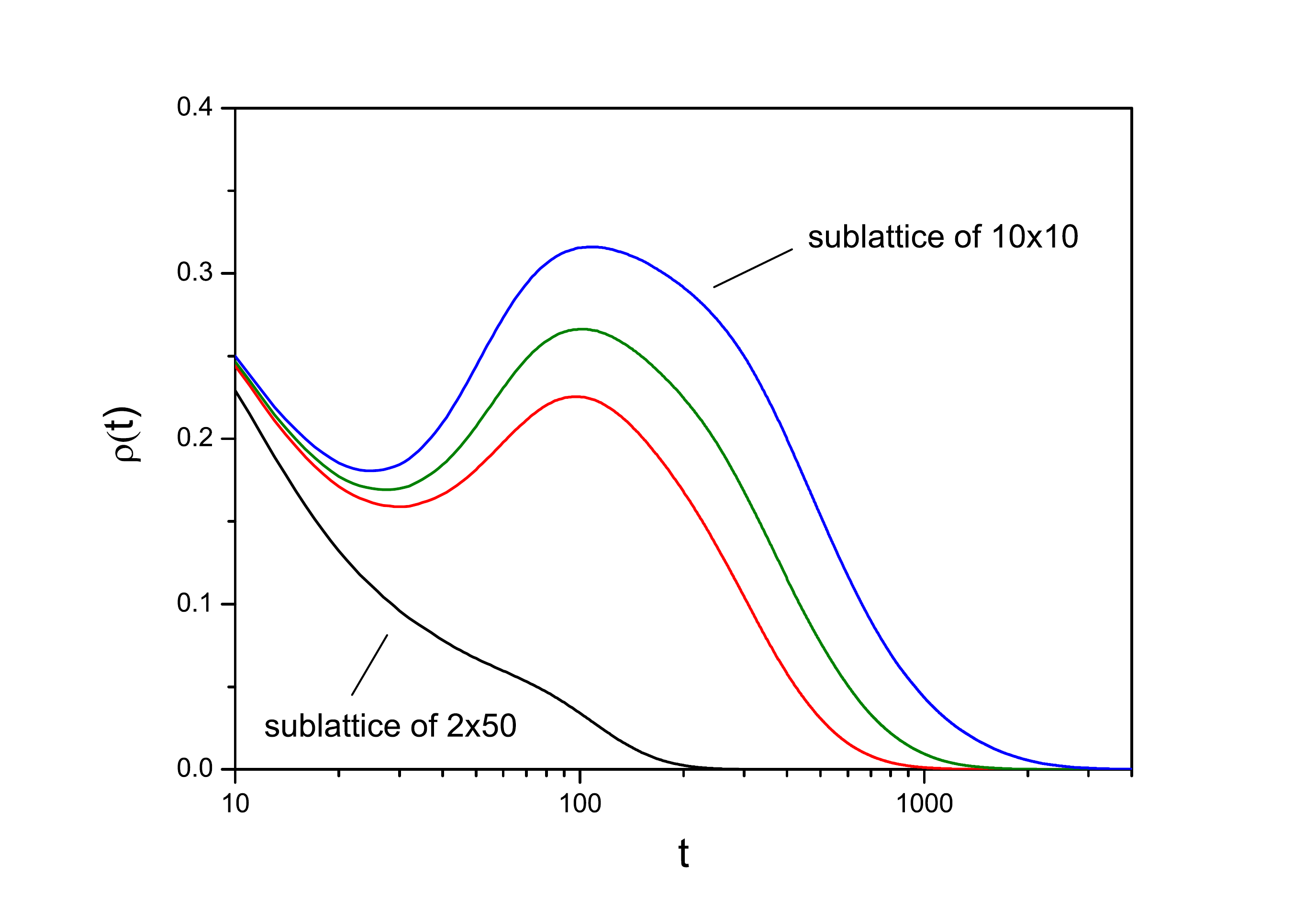}
\par\end{centering}

\caption{\label{fig:figure07}Density $\rho$ as a function of time ($t$)
for a lattice size L = $100\times100$ subdivided into sublattices of different 
aspect
ratios. Four geometric sublattice configurations were considered, from bottom 
to top, $2 \times 50$, $4 \times 25$, $5 \times 20$ and $10 \times 10$; all of 
them 100-site in area size.  Initial
density $\rho(0)=0.8$}
\end{figure}

Strictly considering the model, the behavior of populations differ
for each type of sublattice because the movement changes. The 
Figure~\ref{fig:figure07} shows the effect of the width
by the length (perimeter/area) directly affects the behavior of graphics.
The lower the perimeter/area ratio the greater the viability of a
population (1.04; 0.58; 0.5; 0.4).
Each sublattice
edge will offer different possibilities of internal displacement according
to the position of the individual. In $02\times50$ sublattices there are fewer
opportunities for displacement: there are at most (05) five shift positions
inside the sublattice and up (05) five positions of choice for dispersion
considering that this individual is positioned in top or bottom edge
of its territory. In addition, if a female chooses a partner, but has no 
available
space to allocate their offspring, the newborn dies. At the other
extreme, to a sublattice $10\times10$ the edge effects practically do not exist,
since the possibilities of displacement and dispersion are virtually
total (08 possibilities for movement and dispersion) for any position
of the sublattice. In this case, the behavior will be dictated by total
area and not by perimeter ratio. Since it remains a small lattice,
the population tends to disappear in a relatively short time interval.

\begin{figure}
\begin{centering}
\includegraphics[width=0.9\columnwidth]{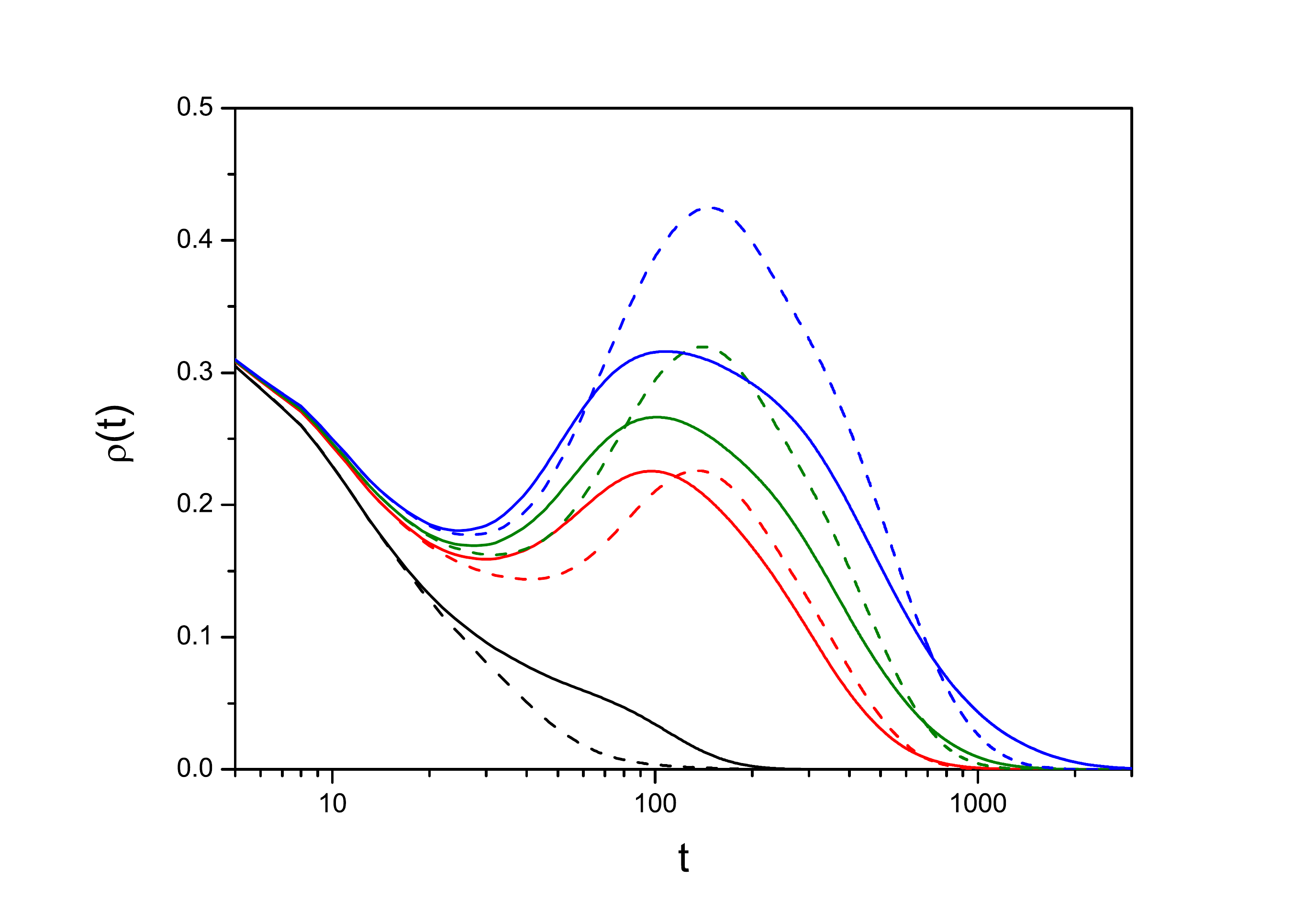}
\par\end{centering}

\caption{\label{fig:figure08}Density $\rho$ as a function of time for a lattice 
size L = $100\times100$ subdivided into 100 sublattices of different aspect 
ratios of 100 site. 
Dashed lines represent the cases where migration was allowed ,  $\phi$ = 0.2 . 
Continuous lines show the simulations without migration. Four geometric 
sublattice configurations were considered, from bottom to top and  both for 
continuous line and dashed line, $02\times50$, $04\times25$, $5\times20$, 
$10\times10$. Initial
density $\rho(0)=0.8$}
\end{figure}

The dispersion is expected to be capable of preventing the extinction
of populations by working as a mechanism of genetic exchange and maintenance
\citep{Hanski2001}. However, their demographic behavior when subject
to an active dispersal in 100-site sublattices, was decreasing until the
total disappearance of the population in all sublattices. Although one
can observe a period of population growth until reaching a maximum
value for sublattices of L = $05\times20$, and $10\times10$, the density 
increase does
not influence survival of populations, possibly due to the effect
of small populations (Figure~\ref{fig:figure08}).

Thus, the contribution of ``identification genomes'' brought by
dispersing individuals among small populations cannot cause changes
that would bring the survival of population. Instead, the dispersion
leads to a curve behavior with a sharper drop. The model does not
consider the continuous flow between fragments, that is, a meta-population
behavior. That would even the loss of balance between individuals
and maintain patches as a single functional 
population~\citep{Hanski-Gaggiotti2004}. Thus, the dispersion between small
lattices, in this model, represents a loss of isolated individuals,
since there is no guarantee that the individual will emigrate to a
lattice minimally populated.

\begin{figure}
\begin{centering}
\includegraphics[width=0.9\columnwidth]{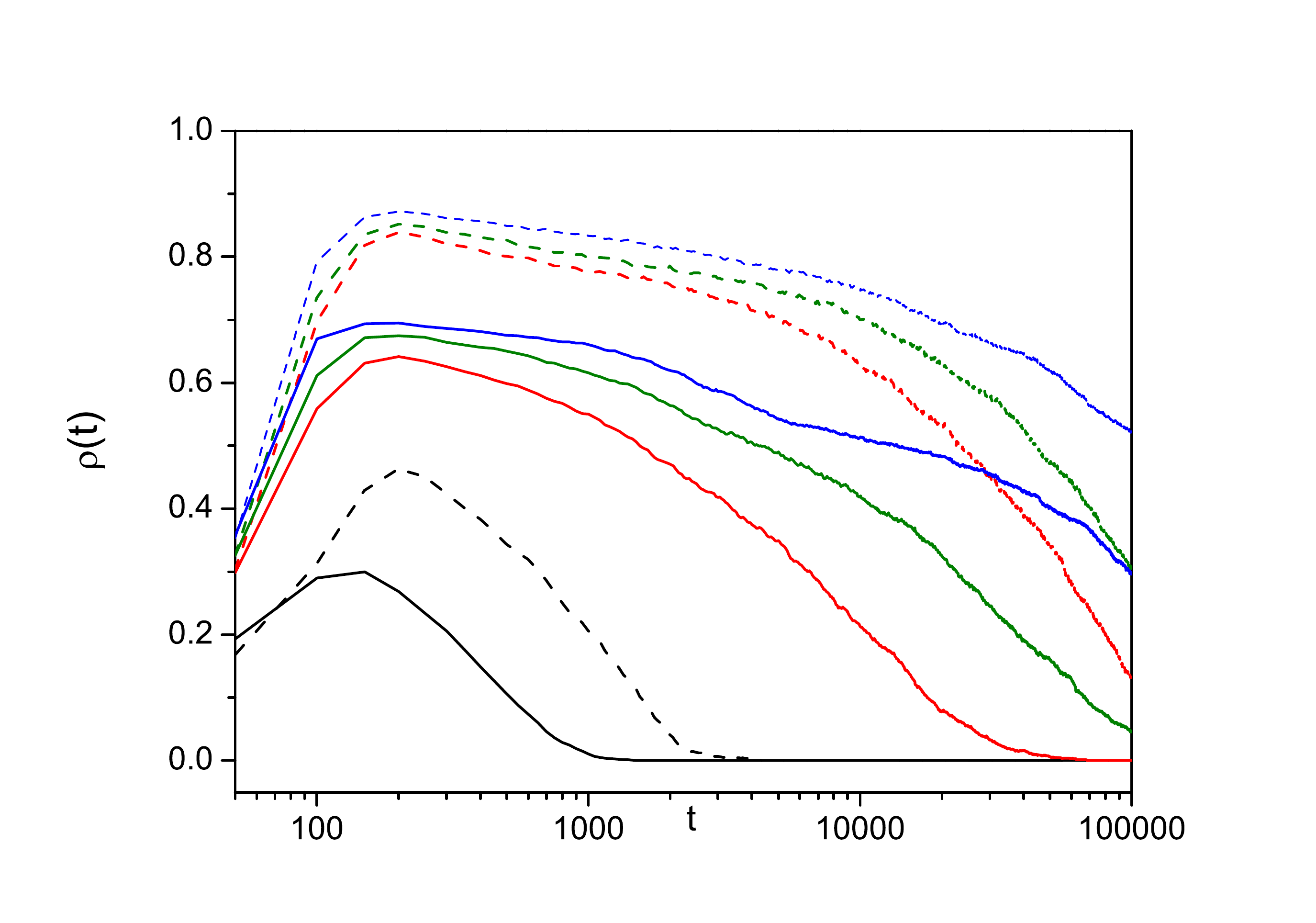}
\par\end{centering}

\caption{\label{fig:figure09}Density $\rho$ as a function of time for a lattice 
size L =$200\times200$ subdivided into 100 sublattices of different aspect 
ratios of 400 site. Dashed lines represent the cases where migration was 
allowed ,  $\phi$ = 0.2 . Continuous lines show the simulations without 
migration. Four geometric sublattice configurations  were considered, from 
bottom to top and  both for continuous line and dashed line, $4\times100$,  
$8\times50$, $10\times40$, $20\times20$.  Initial density $\rho(0)=0.8$.}
\end{figure}

For $200 \times 200$ lattices with sub-lattices of 400 sites, the same behavior
found for sublattices of 100 sites is observed, on a much longer time,
however (Figure~\ref{fig:figure09}). That shows the influence that the area 
plays in maintaining
a population. By quadruplicating
the area, considering the configuration of the sublattice of L = $02\times50$
to 100 sites and L = $04\times100$ to 400 sites, sublattice that would have 
similar
geometric configuration, the average survival increases by a factor
of six.

By introducing the dispersion model for sublattices with L = 400 sites,
the observed demographic behavior completely changes (Figure~\ref{fig:figure09})
leading to believe that the effect of dispersion only makes sense
when related to a (critical) minimum population size. The minimum
value relates to the inflow and outflow of individuals, i.e. the model
indicates there must be a balance between migrants and immigrants
to the variability of ``identification genomes''
effectively raise survival chances.

In natural environments, the minimum population amount to be benefited
with different populations of alleles inputs may not be related to
an absolute value, as noted in the model, but an ecological value,
such as a balance between active adults. Often, a few immigrants should
be able restore the genetic diversity\citep{Mills1996}.

Figure~\ref{fig:figure08} shows a high rate of population
growth, after selecting the ``chronological genome'',
until reaching a maximum value (for L  = $05\times 20$ and $10\times10$ 
lattices),
momentarily reducing the effect of accumulation of deleterious mutations.
This maximum density may represent an ecological value that restores
genetic diversity in nature. However, as the limiting factor in the
model is the space, the population has a higher rate of encounter
(mating) between relatives in small lattices. That quickly eliminates
the effect of variability, increasing the accumulation of deleterious
mutations and leading to population decline. In nature,
the decrease is not due to to genetic effects, but the effects of
lack of resources since the populations are isolated in small areas.

Thus, as observed in the model, in order to achieve survival, ie,
demographic recovery, there must be resources available (sites, in
the model). The entry of individuals in isolated populations can recover
genetic diversity, but can not lead to population survival. Thus,
even if dispersion is present, the recovery of populations in fragmented
landscapes must come associated with measures to ensure the connectivity
of the landscape, such as the formation of ecological corridors that
increase the effective area; also the recovery of degraded areas,
with reforestation involving native species and increase of food items.

\begin{figure}
\begin{centering}
\includegraphics[width=0.9\columnwidth]{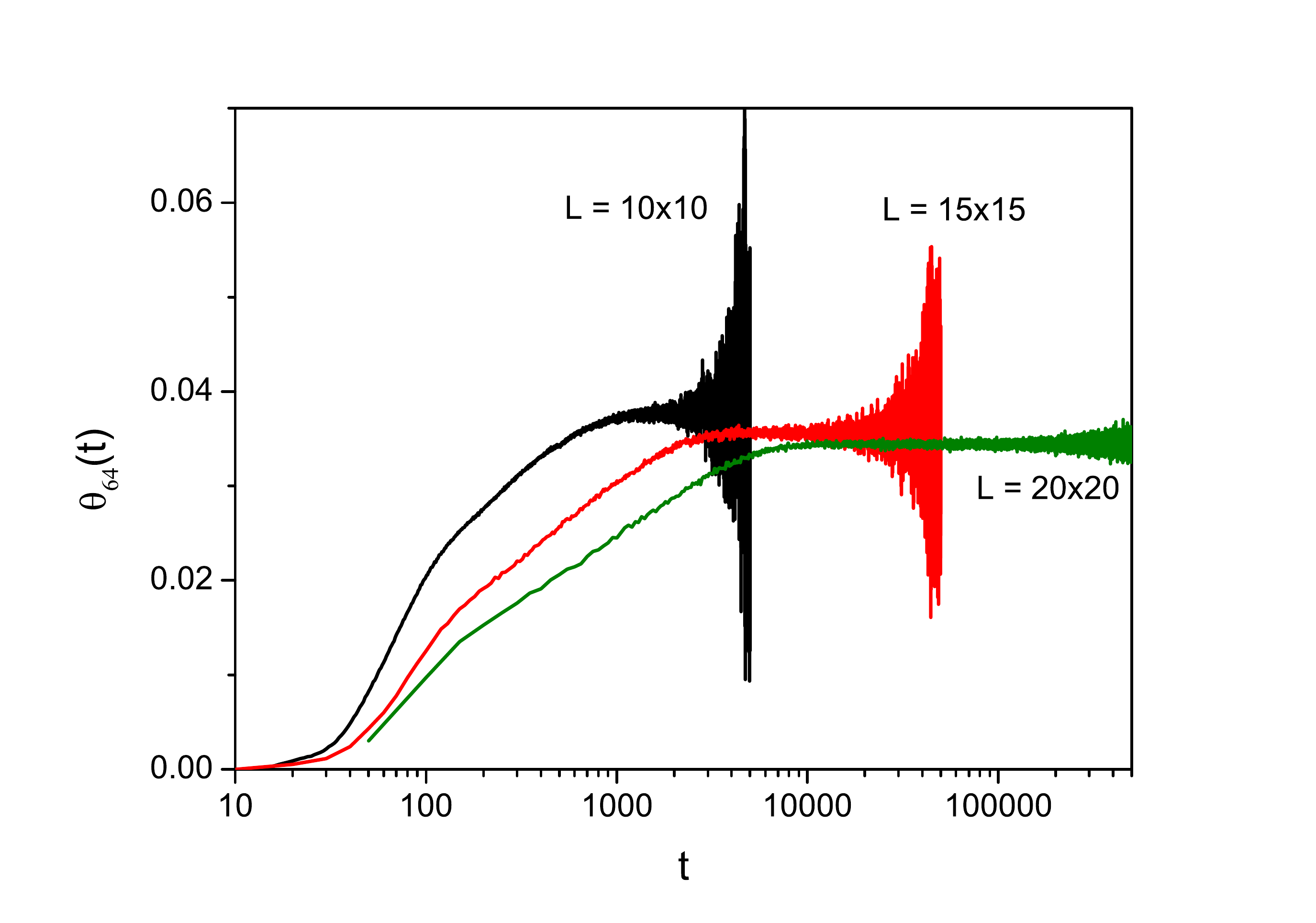}
\par\end{centering}

\caption{\label{fig:figure10}Chronological genome variability for different area
sizes. The time evolution of the fraction $\theta_{64}$ of mating
between relatives with identical pairs of identification genomes is
here shown for lattices of sizes ${10 \times 10}$, $15 \times 15$ and $20 
\times 20$. Large oscillations
can be seen when population becomes so small that statistical deviations
get rather large.  Initial density $\rho(0)=0.8$ and $\phi=0$}
\end{figure}

Figure~\ref{fig:figure10} makes clear the effect of the area
relative to the size of the population, at a fixed density in the
variability of ``identification genomes''
in the population. Once the density is fixed, the larger the lattice is,
the higher will be the number of individuals and therefore the smaller
the fraction $\theta_{64}$ of matings between individuals of identical
pairs of identification genomes. Over time the amount tends to reach
a plateau, given the behavior of the model, i.e a function of predicted
accumulation of deleterious mutations. In the final steps of time,
the oscillatory behavior (high variance) of $\theta_{64}$ is a function
of the density or its absence, down to the point where population
vanishes. To obtain increased survival from individuals flow, the minimum
area and the geometric conformation of the fragments must be taken
into account. In nature, immigration is related to the distance between
fragments and the type of matrix found in the surroundings. Therefore,
no neighborhood effect (availability of sites) is present. In this
model, this effect works as the support capacity of the fragment,
so that fragments with high perimeter/area ratios present little core
area and are unable to provide suitable amounts of resources.

\begin{figure}
\begin{centering}
\includegraphics[width=0.9\columnwidth]{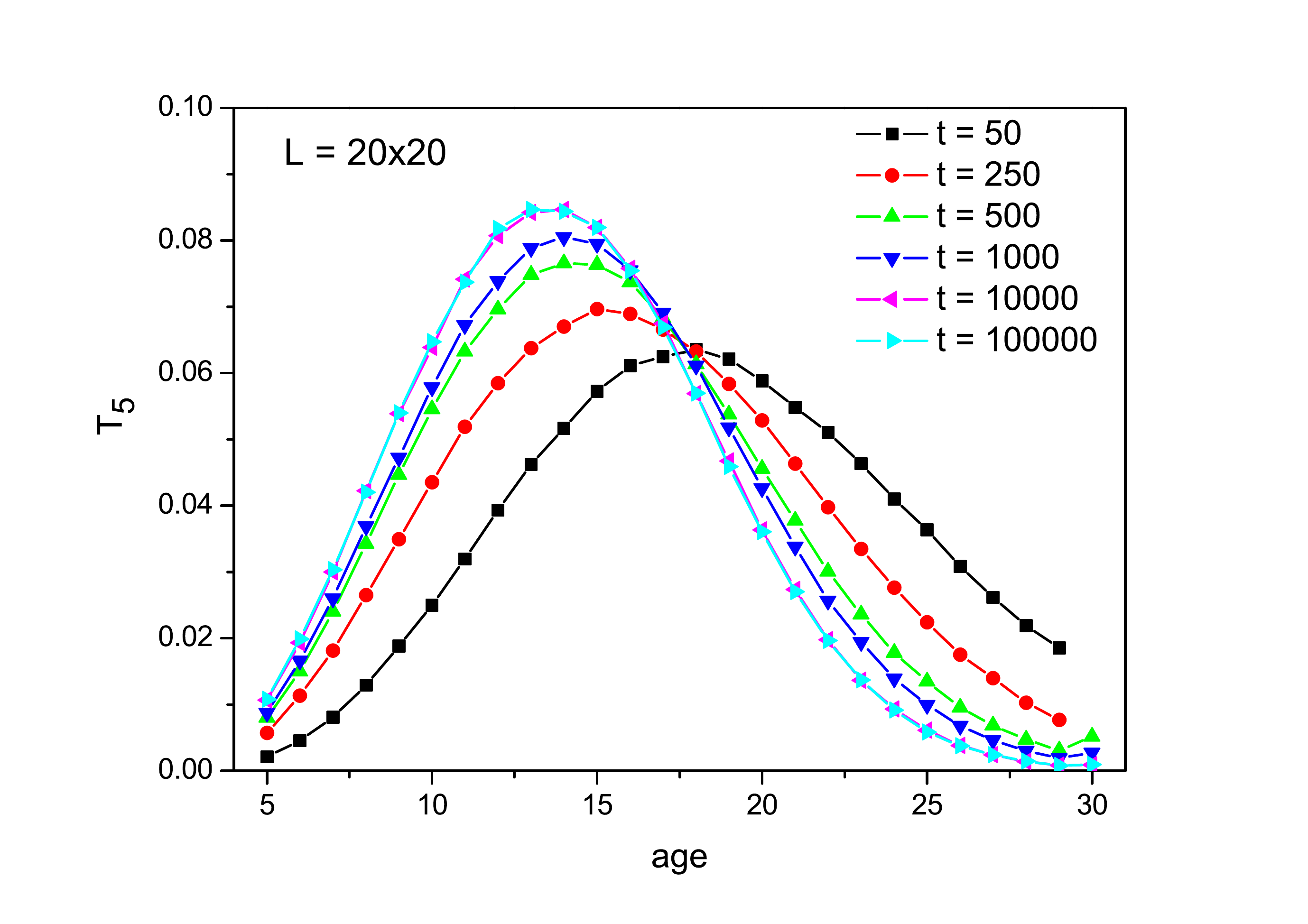}
\par\end{centering}

\caption{\label{fig:figure11}Fraction of population that reach the threshold of 
$T=5$ structured by age, in a lattice 
of $20 \times 20$ at different time periods. }
\end{figure}

\begin{figure}
\begin{centering}
\includegraphics[width=0.9\columnwidth]{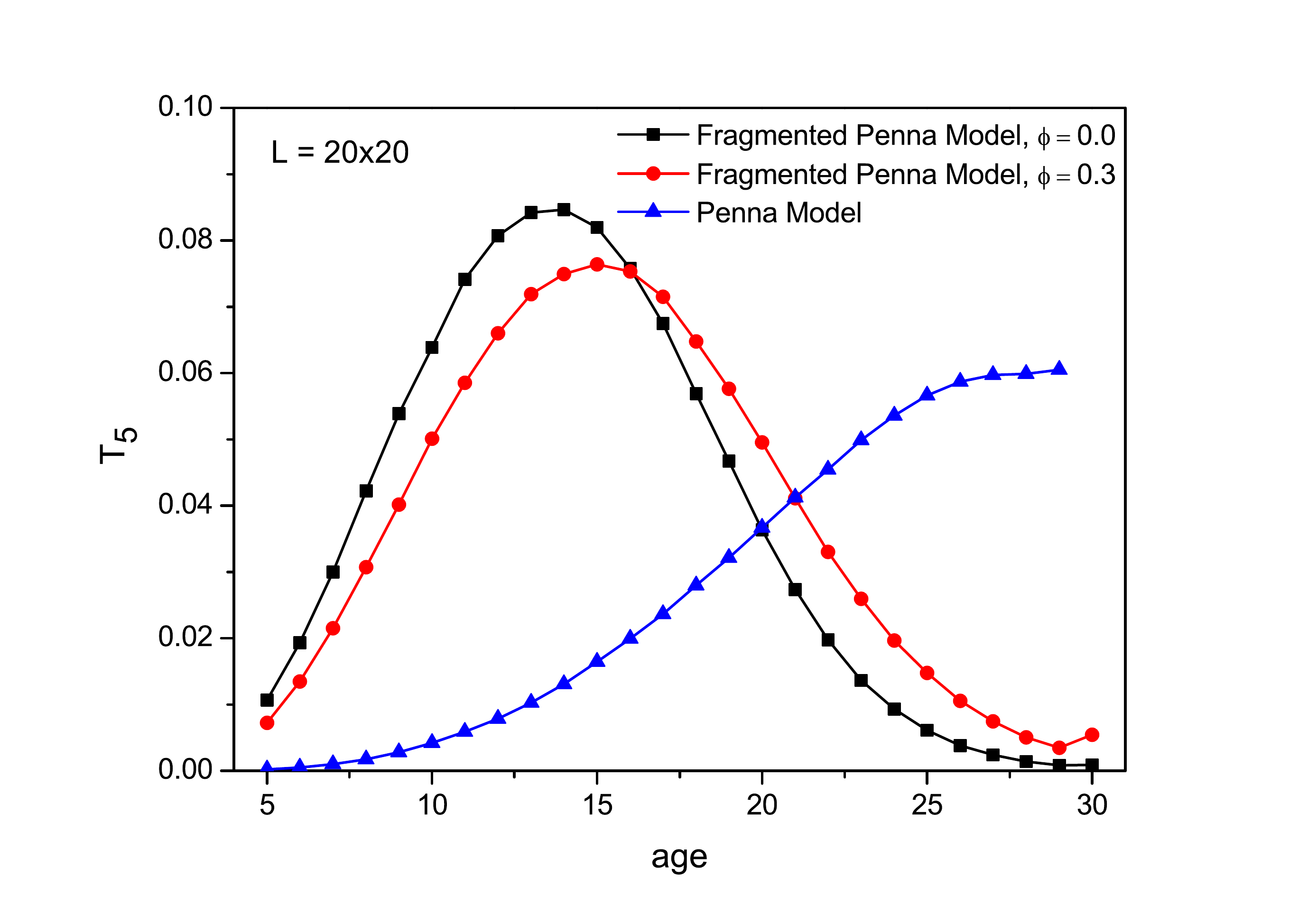}
\par\end{centering}

\caption{\label{fig:figure12} Fraction of population that reach the threshold 
of $T=5$ structured by age, in a lattice 
of $20 \times 20$ and $t=50000$ for Penna Model (triangles), Fragmented Penna 
Model with no dispersion (squares) and Fragmented Penna Model with 
dispersion rate of $\phi=0.3$ (circles). }
\end{figure}

In order to analyze the accumulation of deleterious mutations it was plotted 
the fraction of population that reach the threshold T = 5, ($T_5$)\footnote{It 
is important 
here to note that, once all individuals are diploid, 5 deletery mutations does 
not imply in 5 expressed diseases, due effects of dominance. That way, in the 
text, we refer to threshold T=5 or 
$T_5$, when the individual suffers 5 acummulative diseases.}as a function of 
age; i.e. at what age the individuals are doomed to die, for different times  
(Figure~\ref{fig:figure11}). Note that as the time passes,  the 5th expressed 
disease occurs to younger ages due to the effect of inbreeding in the 
population. It is know that inbreeding is more likely when smaller is the 
population, and fragmented landscapes tends to decrease the population size 
once the individuals are trapped in a 
limited area. 
Figure~\ref{fig:figure12} shows the same data of previous plot ($T_5$ as 
function of age), but now comparing the fragmented Penna model with $\phi = 0$ 
and $\phi=0.30$, and the Penna model with the same paremeters, for the time 
50000, at which the age structure of the population is already stable. It is 
important to remind that, in Figure~\ref{fig:figure12}, the Penna model 
runs on the lattice as the fragmented one, but lacking the part regarding to 
the ``identification genome''.
In Penna model, the 5th expressed disease will occur for older ages, that is, 
the model tends to select individuals over time so that the expressed diseases 
concentrate in more advanced ages.
Differently from what happens with the Penna model, in our model the highest 
incidence of the 5th expressed disease occurs for intermediate ages. And when 
dispersion between networks is added, the effect is slightly minimized, 
with the highest incidence occurring for ages a little older. Note also that 
the curve with the spread after 16 years is always higher than the case without 
dispersion. And the curve without diffusion is higher compared with diffusion 
case, for ages under 16. This shows that diffusion is positive, postponing the 
incidence of the 5th expressed disease.

\subsection{Preservation: Effect of dispersion and variability of 
``chronological
genome'' (gene flow) for females}

\begin{figure}
\begin{centering}
\includegraphics[width=0.9\columnwidth]{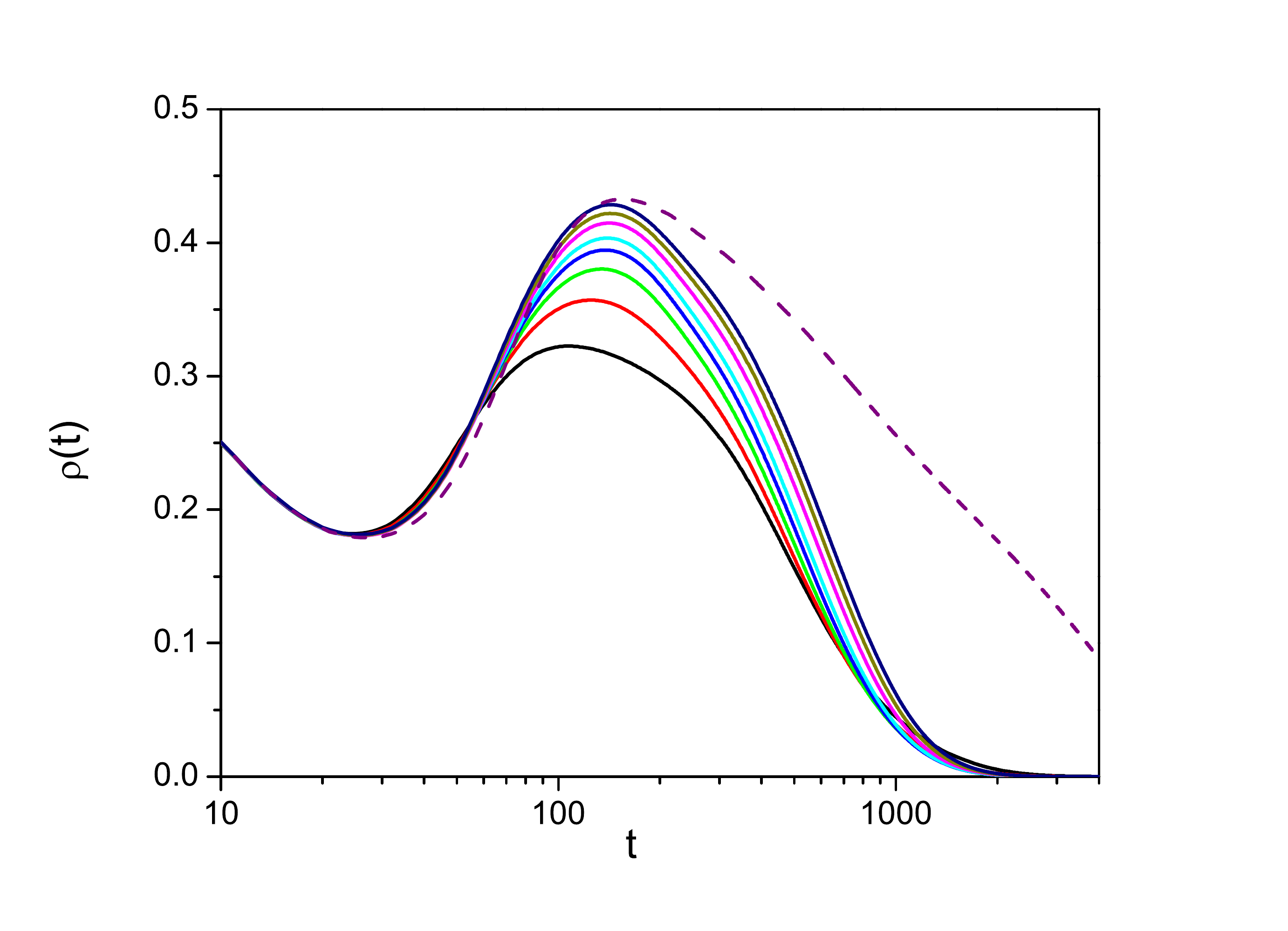}
\par\end{centering}

\caption{\label{fig:figure13}Occupation density $\rho$ as a function of time 
for a $20 \times 20$ square
lattice split into four $10 \times 10$ sub-lattices. The effects of different
dispersion coefficients for males were evaluated, the full lines from bottom to 
top are $\phi$ = 0.0, 0.1, 0.2, 0.3, 0.4, 0.6, 0.8 and 1.0. The special case 
where the migration of females is allowed, $\phi$ = 0.2, instead of males, was 
also evaluated, leading to significantly different results in survival(dashed 
line).}
\end{figure}

The results found showed
a completely different behavior from that observed previously for
males, allowing a considerable survival for the population. When migration
of females was imposed, the expected behavior of the model for individuals'
extinction by the accumulation of deleterious mutations, was shifted
to much longer time steps (Figure~\ref{fig:figure13}).

This effect may have been caused by the increased breeding chances
of the immigrant individual, since it is the female who chooses to
which partner available in her neighborhood she will mate. By moving
from one sublattice to another and meeting females, the chances of success
of an immigrant male is 1:8 to be chosen by females of the new group.
These chances increase now to 7:8 if a female immigrant meets male
neighbors. A male can be surrounded by females, but may simply not
be chosen. With the dispersion of females, the reproductive success,
at every time step, inserts new individuals in the population that
can also reproduce and gradually increase the variability in the population.

It was expected at first that the rate of variability and density
would be more effective with the dispersion of males because males
may mate with different females in the same time step. However, for
a small lattice breed with several females means generating many brothers
and sisters with few sites for displacement, which ends up increasing
the chances of meeting and mating between siblings. In this way, the
effect of deleterious mutations accumulation will overcome
the effect of variability of an immigrant in the population.

Commonly a male can mate with more than one female in a single reproductive
event, subject to the existing hierarchical relations for capuchin
monkeys: multi-male, multi-female system \citep{Fragaszy2004}. In
the present model, this effect is controlled by female choices. In
nature the failure of male immigrants can be brought about by species
behavioral effects, such as male acceptance in the new group, by males
and females, where these choose which and how many males they want
to mate \citep{Fragaszy2004}. In addition, the capuchin monkey females
are said promiscuous, with the alpha male being given priority over
matings, although every one may mate \citep{LynchAlfaro2005,MonicaCarosi2012}.
Thus, copulation is not a guarantee of reproductive success and the
male can take more than a reproductive cycle to pass on their genes
to future generations.

\section{Conclusion and Future Perspectives}

This model was able to describe changes in viability and population dynamics of 
the species crested capuchin considering 
critical levels of survival in a fragmented environment. The effects generated 
by the amount, shape and isolation degree 
of habitat patches available; dispersal ability of species in different 
matrices; and effective population size, 
represented in the model by population density. The simulations results had a 
refined correspondence with what is expected 
to be found in the real space, enabling conservation implications. May be cited:
1) The effects of density (initial density) and lattice size (size of the 
available area) reflects directly in viability of the population.
2) The disappearance of the whole population, viable even for trellis size, is 
a consequence of deleterious mutations accumulation, inherent to the original 
model.
 An analogy to what is found in the effective size of a population.
3) The lower the perimeter/area ratio is; the greater is the viability of a 
population.
4) Dispersion between small lattices, in this model, represents a loss of 
isolated individuals.
5) In order to achieve survival (demographic recovery); there must be resources 
available, represented in the model by empty sites.
6) Even if the dispersion is observed, the recovery of populations in 
fragmented landscapes should come associated with measures to ensure the 
increase of the area for the species.
7) The increase in population survival, even with individuals’ dispersion, 
depends on the minimum area and the geometric conformation of the fragments.
8) Migration of females affects the behavior of the individuals’ extinction by 
the accumulation of deleterious mutations, more than in the case of males’ 
displacement.
 
Finally, it is believed that the overlap of a lattice containing attributes of 
the geographic space in this model (considering potential locations of species 
occurrence) is able to generate future extrapolations, with more reliable 
demographic values, in order to establish the species protection strategies.

\section*{Acknowledgements}

We kindly thank L. Scoss and S. P. Ribeiro for
invaluable discussions and Caio Martim Costa for English manuscript review. 
This work was supported by the Brazilian agencies
FAPEMIG (Fundação de Amparo à Pesquisa do Estado de Minas Gerais),
CAPES (Coordenação de Aperfeiçoamento de Pessoal de Nível Superior)
and CNPq (Conselho Nacional de Desenvolvimento Científico e Tecnológico).




\section*{-----------------}


\bibliographystyle{elsarticle-harv}
\bibliography{library3}

\begin{thebibliography}{75}
\expandafter\ifx\csname natexlab\endcsname\relax\def\natexlab#1{#1}\fi
\expandafter\ifx\csname url\endcsname\relax
  \def\url#1{\texttt{#1}}\fi
\expandafter\ifx\csname urlprefix\endcsname\relax\def\urlprefix{URL }\fi

\bibitem[{Alfaro(2005)}]{LynchAlfaro2005}
Alfaro, J. W.~L., 2005. Male mating strategies and reproductive constraints in
  a group of wild tufted capuchin monkeys (cebus apella nigritus). American
  Journal of Primatology 67~(3), 313--328.

\bibitem[{Alfaro et~al.(2012)Alfaro, Silva, and Rylands}]{Alfaro2012}
Alfaro, J. W.~L., Silva, J. d. S.~E., Rylands, A.~B., 2012. How different are
  robust and gracile capuchin monkeys, an argument for the use of sapajus and
  cebus. American Journal of Primatology 74~(4), 273--286.

\bibitem[{Allee(1927)}]{Allee1927}
Allee, W.~C., 1927. {Animal aggregation}. The Quarterly Review of Biology,
  367--398.

\bibitem[{Allendorf et~al.(2013)Allendorf, Luikart, and Aitken}]{ALLENDORF2013}
Allendorf, F.~W., Luikart, G., Aitken, S.~N., 2013. Conservation and the
  genetics of populations, 2nd edition. Wiley-Blackwell.

\bibitem[{Bennett(2009)}]{Bennett2009}
Bennett, B.~C., 2009. Primate body size – home range relationships ; a
  comparison between four locomotive techniques. Undergraduate Journal of
  Anthropology 1, 131--140.

\bibitem[{Bernardes(1996)}]{BernardesAT1}
Bernardes, A.~T., 1996. Strategies for reproduction and ageing. Annalen der
  Physik 508~(6), 539--549.

\bibitem[{Bernardes et~al.(1998)Bernardes, Moreira, and Castro~e
  Silva}]{bernardes1998simulation}
Bernardes, A.~T., Moreira, J.-G., Castro~e Silva, A., 1998. Simulation of
  chaotic behaviour in population dynamics. The European Physical Journal
  B-Condensed Matter and Complex Systems 1~(3), 393--396.

\bibitem[{Bonte et~al.(2004)Bonte, Lens, and Maelfait}]{Bonte2004}
Bonte, D., Lens, L., Maelfait, J.-P., 2004. Lack of homeward orientation and
  increased mobility result in high emigration rates from low-quality fragments
  in a dune wolf spider. Journal of Animal Ecology 73~(4), 643--650.

\bibitem[{Boyle and Smith(2010)}]{Boyle2010a}
Boyle, S.~A., Smith, A.~T., 2010. Can landscape and species characteristics
  predict primate presence in forest fragments in the brazilian amazon?
  Biological Conservation 143~(5), 1134--1143.

\bibitem[{{Carosi, M. Linn, G. Visalberghi}(2005)}]{MonicaCarosi2012}
{Carosi, M. Linn, G. Visalberghi}, E., 2005. {The Sexual Behavior and Breeding
  System of Tufted Capuchin Monkeys (Cebus apella)}. Advances in the Study of
  Behavior 35, 105--149.

\bibitem[{Castro~e Silva and Bernardes(2001)}]{castro2001analysis}
Castro~e Silva, A., Bernardes, A.~T., 2001. Analysis of chaotic behaviour in
  the population dynamics. Physica A: Statistical Mechanics and its
  Applications 301~(1), 63--70.

\bibitem[{Caughley(1994)}]{Society1994}
Caughley, G., 1994. {Directions in conservation biology}. Journal of Animal
  Ecology 63~(2), 215--244.

\bibitem[{Chaves et~al.(2012)Chaves, Stoner, and
  Arroyo-Rodr{\'\i}guez}]{Chaves2012a}
Chaves, O.~M., Stoner, K.~E., Arroyo-Rodr{\'\i}guez, V., 2012. Differences in
  diet between spider monkey groups living in forest fragments and continuous
  forest in mexico. Biotropica 44~(1), 105--113.

\bibitem[{Chiarello and de~Melo(2001)}]{Chiarello2001}
Chiarello, A., de~Melo, F., 2001. {Primate Population Densities and Sizes in
  Atlantic Forest Remnants of Northern Esp\'{i}rito Santo, Brazil}.
  International Journal of Primatology 22~(3), 379--396.

\bibitem[{Crist{\'o}bal-Azkarate and Arroyo-Rodr{\'\i}guez(2007)}]{Tuxtlas2007}
Crist{\'o}bal-Azkarate, J., Arroyo-Rodr{\'\i}guez, V., 2007. Diet and activity
  pattern of howler monkeys (alouatta palliata) in los tuxtlas, mexico: effects
  of habitat fragmentation and implications for conservation. American Journal
  of Primatology 69~(9), 1013--1029.

\bibitem[{Crooks(2002)}]{Crooks2002}
Crooks, K.~R., 2002. Relative sensitivities of mammalian carnivores to habitat
  fragmentation. Conservation Biology 16~(2), 488--502.

\bibitem[{Dale et~al.(1996)Dale, Pearson, Offerman, and O'Neill}]{Dale1994}
Dale, V., Pearson, S., Offerman, H., O'Neill, R., 1996. Relating patterns of
  land-use change to faunal biodiversity in the central amazon. Biological
  Conservation 2~(76), 216.

\bibitem[{De~Oliveira et~al.(2008)De~Oliveira, Martins, and
  Zacarias}]{DeOliveira2008}
De~Oliveira, A., Martins, S., Zacarias, M., 2008. Computer simulation of the
  coffee leaf miner using sexual penna aging model. Physica A: Statistical
  Mechanics and its Applications 387~(2), 476--484.

\bibitem[{de~Souza et~al.(2009)de~Souza, Martins, and
  Zacarias}]{deSouza2009756}
de~Souza, A., Martins, S., Zacarias, M., 2009. Computer simulation applied to
  the biological control of the insect aphis gossypii for the parasitoid
  lysiphlebus testaceipes. Ecological Modelling 220~(6), 756--763.

\bibitem[{Di~Bitetti and Janson(2001)}]{DiBitetti2001}
Di~Bitetti, M.~S., Janson, C.~H., 2001. Reproductive socioecology of tufted
  capuchins (cebus apella nigritus) in northeastern argentina. International
  Journal of Primatology 22~(2), 127--142.

\bibitem[{Ewers and Didham(2006)}]{Ewers2006}
Ewers, R.~M., Didham, R.~K., 2006. Confounding factors in the detection of
  species responses to habitat fragmentation. Biological Reviews 81~(01),
  117--142.

\bibitem[{Fagan et~al.(1999)Fagan, Cantrell, and Cosner}]{Fagan1999a}
Fagan, W.~F., Cantrell, R.~S., Cosner, C., 1999. How habitat edges change
  species interactions. The American Naturalist 153~(2), 165--182.

\bibitem[{Fahrig(2003)}]{Fahrig2003}
Fahrig, L., 2003. Effects of habitat fragmentation on biodiversity. Annual
  review of ecology, evolution, and systematics, 487--515.

\bibitem[{Ferraz et~al.(2007)Ferraz, Nichols, Hines, Stouffer, Bierregaard, and
  Lovejoy}]{Ferraz2007}
Ferraz, G., Nichols, J.~D., Hines, J.~E., Stouffer, P.~C., Bierregaard, R.~O.,
  Lovejoy, T.~E., 2007. A large-scale deforestation experiment: effects of
  patch area and isolation on amazon birds. science 315~(5809), 238--241.

\bibitem[{Fletcher et~al.(2007)Fletcher, Ries, Battin, and
  Chalfoun}]{Fletcher2007a}
Fletcher, Jr, R.~J., Ries, L., Battin, J., Chalfoun, A.~D., 2007. The role of
  habitat area and edge in fragmented landscapes: definitively distinct or
  inevitably intertwined? Canadian journal of zoology 85~(10), 1017--1030.

\bibitem[{Fragaszy et~al.(2004)Fragaszy, Visalberghi, and
  Fedigan}]{Fragaszy2004}
Fragaszy, D.~M., Visalberghi, E., Fedigan, L.~M., 2004. {The Complete Capuchin:
  The Biology of the Genus Cebus}. Cambridge University Press.

\bibitem[{Gascoigne and Lipcius(2004)}]{Gascoigne2004}
Gascoigne, J.~C., Lipcius, R.~N., 2004. Allee effects driven by predation.
  Journal of Applied Ecology 41~(5), 801--810.

\bibitem[{Gascon et~al.(1999)Gascon, Lovejoy, Jr., Malcolm, Stouffer,
  Vasconcelos, Laurance, Zimmerman, Tocher, and Borges}]{Gascon1999}
Gascon, C., Lovejoy, T.~E., Jr., R. O.~B., Malcolm, J.~R., Stouffer, P.~C.,
  Vasconcelos, H.~L., Laurance, W.~F., Zimmerman, B., Tocher, M., Borges, S.,
  1999. Matrix habitat and species richness in tropical forest remnants.
  Biological Conservation 91~(2--3), 223--229.

\bibitem[{Gehring and Swihart(2003)}]{Gehring2003}
Gehring, T.~M., Swihart, R.~K., 2003. Body size, niche breadth, and
  ecologically scaled responses to habitat fragmentation: mammalian predators
  in an agricultural landscape. Biological Conservation 109~(2), 283--295.

\bibitem[{Gerber et~al.(2012)Gerber, Karpanty, and
  Randrianantenaina}]{Gerber2012}
Gerber, B.~D., Karpanty, S.~M., Randrianantenaina, J., 2012. The impact of
  forest logging and fragmentation on carnivore species composition, density
  and occupancy in madagascar's rainforests. Oryx 46~(03), 414--422.

\bibitem[{Gibson et~al.(2011)Gibson, Lee, Koh, Brook, Gardner, Barlow, Peres,
  Bradshaw, Laurance, Lovejoy, and Sodhi}]{Gibson2011}
Gibson, L., Lee, T.~M., Koh, L.~P., Brook, B.~W., Gardner, T., Barlow, J.,
  Peres, C., Bradshaw, C.~J., Laurance, W.~F., Lovejoy, T.~E., Sodhi, N.~S.,
  2011. Primary forests are irreplaceable for sustaining tropical biodiversity.
  Nature 478~(7369), 378--81.

\bibitem[{Hanski(2001)}]{Hanski2001}
Hanski, I., 2001. Population dynamic consequences of dispersal in local
  populations and in metapopulations. Dispersal. Oxford University Press,
  Oxford, 283--298.

\bibitem[{Hanski and Gaggiotti(2004)}]{Hanski-Gaggiotti2004}
Hanski, I., Gaggiotti, O.~E., 2004. Ecology, genetics, and evolution of
  metapopulations. Academic Press.

\bibitem[{Harcourt and Doherty(2005)}]{HARCOURT2005}
Harcourt, A.~H., Doherty, D.~A., 2005. Species–area relationships of primates
  in tropical forest fragments: a global analysis. Journal of Applied Ecology
  42~(4), 630--637.

\bibitem[{IUCN(2016)}]{IUCN2015}
IUCN, 2016. {Red List of Threatened Species}.

\bibitem[{Janson et~al.(2012)Janson, Baldovino, and {di Bitetti}}]{Janson2012}
Janson, C., Baldovino, M.~C., {di Bitetti}, M., 2012. Long-Term Field Studies
  of Primates. Springer Berlin Heidelberg, Berlin, Heidelberg, Ch. The Group
  Life Cycle and Demography of Brown Capuchin Monkeys (Cebus [apella] nigritus)
  in Iguaz{\'u} National Park, Argentina, pp. 185--212.

\bibitem[{Kim et~al.(2012)Kim, Choi, and Yook}]{kim2012modified}
Kim, Y., Choi, W., Yook, S.-H., 2012. Modified penna bit-string network
  evolution model for scale-free networks with assortative mixing. Journal of
  the Korean Physical Society 60~(4), 621--624.

\bibitem[{Koprowski(2005)}]{Koprowski2005}
Koprowski, J.~L., 2005. {The response of tree squirrels to fragmentation: a
  review and synthesis}. Animal Conservation 8~(4), 369--376.

\bibitem[{Laurance(1997)}]{Laurance1997}
Laurance, W.~F., 1997. {Responses of Mammals to Rainforest Fragmentation in
  Tropical Queensland: a Review and Synthesis}. Wildlife Research 24~(5), 603.

\bibitem[{Laurance et~al.(1998)Laurance, Ferreira, Rankin~de Merona, and
  Laurance}]{Laurance1998}
Laurance, W.~F., Ferreira, L.~V., Rankin~de Merona, J.~M., Laurance, S.~G.,
  1998. Rain forest fragmentation and the dynamics of amazonian tree
  communities. Ecology 79~(6), 2032--2040.

\bibitem[{Laurance et~al.(2002)Laurance, Lovejoy, Vasconcelos, Bruna, Didham,
  Stouffer, Gascon, Bierregaard, Laurance, and Sampaio}]{Laurance2002}
Laurance, W.~F., Lovejoy, T.~E., Vasconcelos, H.~L., Bruna, E.~M., Didham,
  R.~K., Stouffer, P.~C., Gascon, C., Bierregaard, R.~O., Laurance, S.~G.,
  Sampaio, E., 2002. {Ecosystem Decay of Amazonian Forest Fragments: a 22-Year
  Investigation}. Conservation Biology 16~(3), 605--618.

\bibitem[{Lomolino and Perault(2007)}]{Lomolino2007}
Lomolino, M.~V., Perault, D.~R., 2007. Body size variation of mammals in a
  fragmented, temperate rainforest. Conservation Biology 21~(4), 1059--1069.

\bibitem[{Lotka(1925)}]{Lotka1925}
Lotka, A.~J., 1925. Elements of Physical Biology. Williams and Wilkins Company.

\bibitem[{Magdo{\'n} and Maksymowicz(1999)}]{Magdon1999182}
Magdo{\'n}, M.~S., Maksymowicz, A.~Z., 1999. Penna model in migrating
  population--effect of environmental factor and genetics. Physica A:
  Statistical Mechanics and its Applications 273~(1), 182--189.

\bibitem[{Medawar(1952)}]{MEDAWAR1952}
Medawar, P., 1952. An Unsolved Problem of Biology. H.K. Lewis and Company,
  London.

\bibitem[{Mills and Allendorf(1996)}]{Mills1996}
Mills, L.~S., Allendorf, F.~W., 1996. The one-migrant-per-generation rule in
  conservation and management. Conservation Biology 10~(6), 1509--1518.

\bibitem[{Mitani and Rodman(1979)}]{Mitani1979}
Mitani, J.~C., Rodman, P.~S., 1979. {Territoriality: The relation of ranging
  pattern and home range size to defendability, with an analysis of
  territoriality among primate species}. Behavioral Ecology and Sociobiology
  5~(3), 241--251.

\bibitem[{MMA(2014)}]{MMA2014}
MMA, M. d. M.~A., 2014. Portaria nº 444, de 17 de dezembro de 2014. anexo i:
  Lista nacional oficial de espécies da fauna ameaçada de extinção. Diário
  Oficial da União 69~(9), 1013--1029.

\bibitem[{{Moss de Oliveira} et~al.(1999){Moss de Oliveira}, Bernardes, and
  {S\'{a} Martins}}]{MossdeOliveira1999}
{Moss de Oliveira}, S., Bernardes, A.~T., {S\'{a} Martins}, J., 1999.
  {Self-organisation of female menopause in populations with child-care and
  reproductive risk}. The European Physical Journal B 7~(3), 501--504.

\bibitem[{Murcia(1995)}]{Murcia1995}
Murcia, C., 1995. {Edge effects in fragmented forests: implications for
  conservation.} Trends in ecology \& evolution 10~(2), 58--62.

\bibitem[{Oliveira et~al.(2003)Oliveira, Stauffer, and Sousa}]{Oliveira2003}
Oliveira, S. M.~D., Stauffer, D., Sousa, A.~O., 2003. {Computer Simulation of
  Sympatric Speciation with Penna Ageing Model III The Penna model with
  phenotype selection}. Brazilian Journal of Physics 33~(3), 623--627.

\bibitem[{Pearce et~al.(2013)Pearce, Carbone, Cowlishaw, and
  Isaac}]{Pearce2013}
Pearce, F., Carbone, C., Cowlishaw, G., Isaac, N. J.~B., 2013. Space-use
  scaling and home range overlap in primates. Proceedings of the Royal Society
  B: Biological Sciences 280~(1751).

\bibitem[{Penna et~al.(1995)Penna, de~Oliveira, and
  Stauffer}]{PhysRevE.52.R3309}
Penna, T., de~Oliveira, S.~M., Stauffer, D., 1995. Mutation accumulation and
  the catastrophic senescence of the pacific salmon. Physical Review E 52~(4),
  R3309.

\bibitem[{Penna and Stauffer(1995)}]{Penna1995a}
Penna, T., Stauffer, D., 1995. Efficient monte carlo simulation of biological
  aging. International Journal of Modern Physics C 6~(02), 233--239.

\bibitem[{Penna(1995)}]{Penna1995}
Penna, T. J.~J., 1995. {A bit-string model for biological aging}. Journal of
  Statistical Physics 78~(5), 1629--1633.

\bibitem[{Pi{\~n}ol and Banzon(2011)}]{Pinol2011a}
Pi{\~n}ol, C.~M., Banzon, R., 2011. Catastrophic senescence and semelparity in
  the penna aging model. Theory in Biosciences 130~(2), 101--106.

\bibitem[{Pontes et~al.(2012)Pontes, de~Paula, and Magnusson}]{Pontes2012}
Pontes, A. R.~M., de~Paula, M.~D., Magnusson, W.~E., 2012. {Low Primate
  Diversity and Abundance in Northern Amazonia and its Implications for
  Conservation}. Biotropica 44~(6), 834--839.

\bibitem[{P\"{u}ttker et~al.(2013)P\"{u}ttker, Bueno, de~Barros, Sommer, and
  Pardini}]{Puttker2013}
P\"{u}ttker, T., Bueno, A.~A., de~Barros, C. d.~S., Sommer, S., Pardini, R.,
  2013. {Habitat specialization interacts with habitat amount to determine
  dispersal success of rodents in fragmented landscapes}. Journal of Mammalogy
  94~(3), 714--726.

\bibitem[{Rai(2003)}]{Rai2003}
Rai, U.~K., 2003. {Minimum Sizes for Viable Population and Conservation
  Biology}. Nature 1, 3--9.

\bibitem[{Ries and Sisk(2004)}]{Ries2004}
Ries, L., Sisk, T.~D., 2004. A predictive model of edge effects. Ecology
  85~(11), 2917--2926.

\bibitem[{Robinson et~al.(1995)Robinson, Thompson, Donovan, Whitehead, and
  Faabrog}]{Robinson1995}
Robinson, S.~K., Thompson, F.~R., Donovan, T.~M., Whitehead, D.~R., Faabrog,
  J., 1995. Regional forest fragmentation and the nesting success of migratory
  birds. Science 267, 1987--1990.

\bibitem[{Rylands et~al.(2005)Rylands, Kierulff, and Mittermeier}]{rylands2005}
Rylands, A.~B., Kierulff, M. C.~M., Mittermeier, R.~A., 2005. Notes on the
  taxonomy and distributions of the tufted capuchin monkeys (cebus, cebidae) of
  south america. Lundiana 6, 97--110.

\bibitem[{Saunders et~al.(1991)Saunders, Hobbs, and Margules}]{SAUNDERS1991}
Saunders, D.~A., Hobbs, R.~J., Margules, C.~R., 1991. {Biological Consequences
  of Ecosystem Fragmentation: A Review}. Conservation Biology 5~(1), 18--32.

\bibitem[{Shaffer(1981)}]{Shaffer1981}
Shaffer, M.~L., 1981. {Minimum Population Sizes for Species Conservation}.
  BioScience 31~(2), 131--134.

\bibitem[{Silva~Jr.(2001)}]{JUNIOR2001}
Silva~Jr., J. S.~E., 2001. Especia{\c{c}}{\~a}o nos macacos-prego e caiararas,
  g{\^e}nero {Cebus} erxleben, 1777 ({Primates}, {Cebidae}). Ph.D. thesis,
  Universidade Federal do Rio de Janeiro.

\bibitem[{{{SOS Mata Atl\^{a}ntica} and
  INPE}(2013)}]{SOSMataAtlantica;INPE2013}
{{SOS Mata Atl\^{a}ntica} and INPE}, 2013. {Atlas dos remanescentes florestais
  da Mata Atl\^{a}ntica per\'{i}odo 2011-2012}.

\bibitem[{Soul\'{e}(1985)}]{Soule1985}
Soul\'{e}, M.~E., 1985. {What is Conservation Biology?} BioScience 35~(11),
  727--734.

\bibitem[{Sousa et~al.(2000)Sousa, de~Oliveira, and
  Bernardes}]{sousa2000simulating}
Sousa, A. d.~O., de~Oliveira, S.~M., Bernardes, A.~T., 2000. Simulating
  inbreeding depression through the mutation accumulation theory. Physica A:
  Statistical Mechanics and its Applications 278~(3), 563--570.

\bibitem[{Srinivasaiah et~al.(2012)Srinivasaiah, Anand, Vaidyanathan, and
  Sinha}]{Srinivasaiah2012}
Srinivasaiah, N.~M., Anand, V.~D., Vaidyanathan, S., Sinha, A., 2012. {Usual
  populations, unusual individuals: insights into the behavior and management
  of Asian elephants in fragmented landscapes.} PloS one 7~(8), e42571.

\bibitem[{Stauffer et~al.(1996)Stauffer, de~Oliveira, de~Oliveira, and dos
  Santos}]{Stauffer1996}
Stauffer, D., de~Oliveira, P., de~Oliveira, S., dos Santos, R., 1996. {Monte
  Carlo simulations of sexual reproduction}. Physica A: Statistical Mechanics
  and its Applications 231~(4), 504--514.

\bibitem[{{Torres de Assump\c{c}\~{a}o}(1983)}]{Torres-de-assuncao-1983}
{Torres de Assump\c{c}\~{a}o}, C., 1983. An ecological study of the primates of
  south-eastern {Brazil}, with an re-appraisal of {"Cebus apella"} races. Ph.D.
  thesis, Univ. of Edinburgh.

\bibitem[{Vitousek et~al.(1997)Vitousek, Mooney, Lubchenco, and
  Melillo}]{Vitousek-1997}
Vitousek, P.~M., Mooney, H.~A., Lubchenco, J., Melillo, J.~M., 1997. Human
  domination of earth's ecosystems. Science 277~(5325), 494--499.

\bibitem[{Volterra(1926)}]{Volterra1926}
Volterra, V., 1926. {Fluctuations in the abundance of a species considered
  mathematically}. Nature 118, 558--560.

\bibitem[{Weinberg(1908)}]{Gotelli2008}
Weinberg, W., 1908. Über den Nachweis der Vererbung beim Menschen. Vol.~64.

\bibitem[{Young et~al.(2000)Young, Industry, Genetics, Clarke, Entomology,
  Ecology, and Plans}]{Young2000}
Young, A., Industry, C.~P., Genetics, F.~C., Clarke, G., Entomology, C.,
  Ecology, M., Plans, A., 2000. {Genetics , Demography and Viability of
  Fragmented Populations}. Cambridge University Press.

\end{thebibliography}

\end{document}